\documentclass[12pt]{article}
\pdfoutput=1
\usepackage[utf8]{inputenc}
\usepackage[T1]{fontenc}
\usepackage[normalem]{ulem}
\usepackage{verbatim}
\usepackage{amsmath,amssymb,amsthm,amsfonts,mathrsfs,dsfont,braket,array,float,afterpage}
\usepackage{graphicx}
\usepackage[dvipsnames]{xcolor}
\usepackage{slashed}
\usepackage{enumitem}
\usepackage{subfiles}

\usepackage[colorlinks=false,linkcolor=darkblue,  urlcolor=blue,    filecolor=blue,     citecolor=red,
linktocpage=true,pdfstartview=FitV,bookmarksopen=true,
pdfencoding=auto,
pagebackref]{hyperref}

\usepackage{cleveref}
\usepackage{epsfig}
\usepackage{mathabx}
\usepackage[mathscr]{eucal}
\usepackage{stmaryrd}
\usepackage{comment}
\usepackage{esint}
\usepackage{physics}
\usepackage{setspace}
\usepackage{braket}
\usepackage{float}
\usepackage{url}
\usepackage{mathtools}
\usepackage{bbold}
\usepackage{slashed}
\usepackage{tikz}
\usetikzlibrary{decorations.markings}
\usetikzlibrary{decorations.pathmorphing}
\usepackage{adjustbox}
\usepackage{graphicx}
\usepackage{epstopdf}
\usepackage[labelfont=bf]{caption}
\usepackage{subcaption}
\usepackage{pgfplots}
\usepackage{fancybox}
\usepackage{eurosym}
\usepackage{tcolorbox}
\usepackage{tensor}
\usepackage{soul}
\usepackage{comment}
\usetikzlibrary{arrows.meta}  
\allowdisplaybreaks

\usepackage[margin = 2.37cm]{geometry}
\usepackage[ragged]{footmisc}
\definecolor{orangewolfram}{RGB}{224.584, 155.815, 36.223}
\definecolor{bluewolfram}{RGB}{93.9463, 129.229, 180.998}
\definecolor{greenwolfram}{RGB}{142.846, 176.35, 49.6957}

\def\be{\begin{equation}}
	\def\ee{\end{equation}}

\def\Pf{\mathrm{Pf}}

\numberwithin{equation}{section}

\def\cW{{\cal W}}
\def\da{{\dot{a}}}
\def\db{{\dot{b}}}
\def\dc{{\dot{c}}}
\def\dd{{\dot{d}}}
\def\cW{{\cal W}}
\def\cM{{\cal M}}

\usepackage{formatting}

\usepackage{shortcuts}

\newcommand{\drawsquare}[2]{\hbox{%
\rule{#2pt}{#1pt}\hskip-#2pt
\rule{#1pt}{#2pt}\hskip-#1pt
\rule[#1pt]{#1pt}{#2pt}}\rule[#1pt]{#2pt}{#2pt}\hskip-#2pt
\rule{#2pt}{#1pt}}

\newcommand{\fund}{~\raisebox{-.5pt}{\drawsquare{6.5}{0.4}}~}

\pgfplotsset{compat=1.18}

\begin{document}
\begin{titlepage}
\begin{flushright}
SISSA 06/2025/FISI
\end{flushright}
\bigskip
\def\thefootnote{\fnsymbol{footnote}}

\begin{center}
{\LARGE
{\bf Gravity, finite duality cascades and confinement} 
}
\end{center}

\vskip 5pt

\begin{center}
{\large
Fabrizio Aramini$^{a,b}$, 
Riccardo Argurio$^{c}$, 
Matteo Bertolini$^{a,b}$,
\vskip 4pt
Eduardo Garc\'{\i}a-Valdecasas$^{a,b}$, 
Pietro Moroni$^{a,b}$}

\renewcommand{\thefootnote}{\arabic{footnote}}

\vspace{0.25cm}
{\large 
$^a$ {\it SISSA, 
Via Bonomea 265; I 34136 Trieste, Italy\\}
$^b$ {\it INFN - Sezione di Trieste,  
Via Valerio 2, 34127 Trieste, Italy\\}
$^c${\it Physique Th\'eorique et Math\'ematique and International Solvay Institutes \\ 
Universit\'e Libre de Bruxelles, C.P. 231, 1050 Brussels, Belgium}
}
\vskip 5pt
{\texttt{faramini,bertmat,egarciav,pmoroni @sissa.it, Riccardo.Argurio@ulb.be}}

\end{center}

\begin{center} {\bf Abstract} \end{center}
 
\noindent
Cascading RG flows are characteristic of $\mathcal{N}=1$ gauge theories realized by D3-branes probing singularities in the presence of fractional branes. A celebrated example is the Klebanov-Strassler model, which exhibits an infinite cascade that ends with  confinement. In this work, we explore a related setup where the addition of an orientifold plane modifies the cascade structure: the RG flow now consists of a finite number of steps, originating from a UV fixed point with a finite number of degrees of freedom. We provide a supergravity solution dual to this flow, that reproduces all its salient features. The string frame curvature can be kept small up to parametrically large values of the holographic coordinate, larger than the one at which the first cascade step occurs.\let\thefootnote\relax\footnotetext{\em On the occasion of the 25$^{\text \, th}$ anniversary of the Klebanov-Strassler solution.}

\end{titlepage}

\setcounter{footnote}{0}

\newpage
\tableofcontents

\section{Introduction and summary of results}
\label{Intro}

Since the early stages of the AdS/CFT correspondence \cite{Maldacena:1997re,Gubser:1998bc,Witten:1998qj}, significant efforts have been devoted to extending it beyond the original duality between ${\cal N}=4$ SYM and type IIB string theory on $AdS_5 \times S^5$, aiming to include theories with reduced supersymmetry and non-conformal dynamics.

As part of this broader program, a wide class of ${\cal N}=1$ supersymmetric gauge theories have been realized through configurations of D-branes located at Calabi–Yau (CY) singularities. These theories are  typically quiver gauge theories, comprising several unitary gauge groups and matter in bi-fundamental representations.

When all gauge groups possess equal ranks, the resulting theories are superconformal. The dynamics becomes more intricate with the introduction of fractional D-branes, which explicitly break conformal invariance and trigger non-trivial renormalization group (RG) flows in the dual gauge theory (see,  e.g. \cite{Bertolini:2003iv} for a pedagogical introduction).

The canonical example of this setup is the model presented in  \cite{Klebanov:1998hh,Klebanov:2000nc,Klebanov:2000hb}, which is realized by placing a stack of $N$ regular and $M$ fractional D3-branes at the tip of the conifold singularity. For $M=0$, the dual gauge theory has gauge group $SU(N) \times SU(N)$ with bifundamental matter and a quartic superpotential. The theory possesses a two-dimensional complex manifold of superconformal fixed points, which is conjectured to be dual to Type IIB string theory on $AdS_5 \times T^{1,1}$ with $N$ units of $F_5$ flux \cite{Klebanov:1998hh}. 
By adding fractional D3-branes, one obtains an $\mathcal{N}=1$ $SU(N + M) \times SU(N)$ gauge theory with the same matter and superpotential as its superconformal counterpart, though the theory is no longer conformal \cite{Klebanov:2000nc}. The RG flow can be efficiently described in terms of a cascade of Seiberg dualities.\footnote{We refer to the masterpiece \cite{Strassler:2005qs} for a detailed explanation.} The theory possesses a huge moduli space of supersymmetric vacua which, for generic values of $N$ and $M$, is a multidimensional mesonic branch \cite{Dymarsky:2005xt}. Whenever the number of D3-branes is proportional to the number of fractional D3-branes, the theory also admits a baryonic branch of complex dimension one. The corresponding RG flow is described by the Type IIB supergravity solution, due to Klebanov and Strassler (KS), known as the warped deformed conifold \cite{Klebanov:2000hb}.\footnote{The KS solution describes the maximally symmetric point of the baryonic branch. See \cite{Butti:2004pk} for a description of the whole one-dimensional baryonic branch.} 

One key feature of the KS model (and any of its generalizations) is that the (effective) number of regular D3-branes diminishes toward the IR. The number of fractional D3-branes, instead, remains constant and equal to $M$ at any energy scale. The fact that $M$ is constant implies that the duality cascade continues indefinitely toward higher and higher energies. In other words, the theory does not admit a properly defined UV fixed point, either weakly or strongly coupled, the UV completion being in terms of an infinite number of degrees of freedom. In this sense, the field theory is rather peculiar.

As observed in \cite{Argurio:2017upa}, the addition of orientifolds to this setup can drastically change the state of affairs. Unlike in unorientifolded singularities, in presence of orientifolds, more specifically of O7-planes, the number of fractional branes is not preserved along the RG flow.\footnote{This is also a feature of quiver gauge theories with fundamental flavors, see e.g. \cite{Benini:2006hh,Benini:2007gx,Benini:2007kg,Bigazzi:2008ie,Bigazzi:2008qq}.} This has several consequences, the most striking being that duality cascades now become {\it finite}. This implies that, unlike ordinary duality cascades, orientifold cascades can possess UV completions that do not require an infinite number of degrees of freedom.\footnote{See \cite{Chatzis:2024top,Chatzis:2025dnu} for a different class of RG flows from a UV fixed point to confinement, albeit with dimensional reduction along the flow, from four to three space-time dimensions. For an example of a finite cascade entirely in three dimensions see \cite{Faedo:2022lxd}.} In the far UV, the theory enters a (free or interacting) conformal phase after a finite RG time. If the fixed point is interacting, one can then hope to end up with gauge theories that interpolate between a well-defined fixed point in the UV and, for example, confinement in the IR—while maintaining, in principle, a weakly curved gravity dual description throughout the entire RG flow!

In this work, we show that the simple setup consisting of the original conifold theory supplemented with an O7-plane explicitly realizes all of the above ideas. 

The introduction of orientifolds within the gauge/gravity correspondence allows for dual gauge theories that include orthogonal and symplectic gauge groups, in addition to unitary ones \cite{Witten:1998xy}. The conifold admits several supersymmetry-preserving orientifold projections, each giving rise to a different gauge theory. We consider an O7-plane  embedded \`a la Kuperstein \cite{Kuperstein:2004hy} which, as shown in \cite{Naculich:2002pq}, gives rise to a $USp(N+M-2) \times USp(N)$ gauge theory, where $N$ and $M$ are the numbers of regular and fractional D3-branes, respectively.\footnote{Different versions of the gauge theory obtained by placing an O-plane at the conifold singularity were previously discussed in \cite{Imai:2001cq,Naculich:2002pq}. In \cite{Imai:2001cq} this was a $USp \times SO$ theory, that displays an infinite duality cascade as the original KS model \cite{Klebanov:2000hb}. The theory discussed in \cite{Naculich:2002pq} has instead $USp \times USp$ gauge group, as ours, but the addition of flavor D7-branes that was considered there leads again to an RG flow with an infinite cascade. See, e.g., \cite{Franco:2008jc,Franco:2015kfa} for other examples of duality cascades in presence of orientifolds.} 

The RG flow is described by a duality cascade. At each step of the cascade toward the IR, the number of fractional branes increases by 4, while the number of regular branes decreases by $M+2$.\footnote{We use conventions such that $N$ and $M$ are even, and hence correspond to the number of regular and fractional D3-branes in the covering, unorientifolded ten-dimensional spacetime.} This implies that, conversely, $M$ diminishes toward the UV. When $M$ becomes as small as 2 or 4, the (upward) cascade comes to a stop and the theory reaches a one-dimensional conformal manifold. 

As for the IR dynamics and the phases of the theory, the moduli space of supersymmetric vacua consists of a multidimensional mesonic branch. No baryonic branch is expected given that there are no baryonic operators for $USp$ groups.
Consequently, the string dual is expected to correspond to a smooth geometric background populated by mobile D3-branes. An exception to this general result occurs if at the last cascade step in the IR the effective number of regular and fractional branes satisfies $N_{\text{IR}}=M_{\text{IR}}+2$. In this case, the theory also exhibits an isolated vacuum at the origin of field space, where all mesons are massive and can be integrated out. The corresponding low-energy effective theory becomes pure $USp(M_{\text{IR}}+2)$ SYM, which confines and for which one expects a purely geometric dual, very much like the symmetric point of the baryonic branch of the KS model. This is the vacuum whose gravity dual we aim to describe. 

The gravity dual description of the orientifold gauge theory is realized by considering the backreaction of D3-branes placed at the tip of the conifold geometry with an O7-plane. In contrast to the parent KS solution, the presence of the O7-plane leads to a background in which both the RR scalar potential $C_0$ and the dilaton have a non-trivial profile. This is expected, as O-planes behave similarly to D7-branes, as far as coupling to closed string fields is concerned. Therefore, we adopt the same ansatz as in  \cite{Benini:2006hh,Benini:2007gx}, where the flavored version of the KS solution was derived. However, an O7-plane effectively backreacts as D7-branes with {\it negative} tension. This has dramatic consequences. 

For small values of the holographic coordinate, the geometry reduces to the deformed conifold, in agreement with the expected confining behavior of the dual gauge theory. Interestingly, a potential divergence caused by the dilaton profile, that would correspond to a IR duality wall in the dual gauge theory ({\it i.e.} an accumulation point of Seiberg dualities), is cured in a way similar to the way the KS solution \cite{Klebanov:2000hb} resolves the short distance singularity of the KT solution \cite{Klebanov:2000nc}. The solution also correctly captures the duality cascade, with the RR $F_5$ flux decreasing and the $F_3$ flux increasing toward the IR. At large radii, the solution asymptotes to $AdS_5 \times T^{1,1}$ with $N_{\text{UV}}$ units of $F_5$ flux, again in agreement with field theory expectations, where a fixed point is anticipated. One seemingly subtle aspect concerns the dilaton profile, which is such that the effective string coupling vanishes as $r \to \infty$. However,  we show that by appropriately tuning the free parameters on which the bulk solution depends, it is possible to ensure that the string-frame curvature, proportional to an inverse power of $g_s N_{\text{UV}}$, remains sufficiently small over a broad range of energy scales. In particular, it can be made sufficiently small throughout the entire RG flow, from the far UV to the deep IR. This, in principle, renders the dual supergravity solution amenable to performing gauge/gravity duality checks that go beyond those typically accessible in generic non-conformal quiver gauge theories, whose UV completions are less conventional. In this work, we begin by analyzing more traditional checks, such as the correct description of the RG flow in terms of a duality cascade, the computation of the $\beta$-functions, the pattern of R-symmetry breaking, and the holographic calculation of the $a$- and $c$-charges at the UV fixed point. In future work, we plan to explore additional aspects of the duality.

In the remainder of the paper, we provide a detailed exposition supporting our claims. In section \ref{sec:field theory analysis}, we present the gauge theory and analyze several of its key features, including the duality cascade, the structure of the moduli space of confining vacua, and the nature of the UV fixed point. Section \ref{sec: SUGRA solution} introduces the corresponding supergravity solution, with the full derivation provided in Appendix \ref{app: SUGRA solution}. Finally, in section \ref{sec: checks}, we carry out the consistency checks mentioned above, in support of the proposed duality.

\section{The orientifold field theory}
\label{sec:field theory analysis}

The conifold $\mathcal{C}$ is a singular Calabi-Yau $3$-fold that can be described as a complex three-dimensional hypersurface in $\mathbb{C}^4$
defined by the equation
\begin{equation}
\label{eq: conifold}
    z_1^2+ z_2^2 + z_3^2 + z_4^2=0 \ .
    \end{equation}
Consider Type IIB String Theory on $M_{10} =\mathbb{R}^4 \times \mathcal{C}$ in the presence of a stack of (regular and fractional) D3-branes placed at the tip of the conifold and spanning $\mathbb{R}^4$. In the decoupling limit, the low-energy dynamics on the D-branes is described by the Klebanov-Strassler ${\cal N}=1$ quiver gauge theory \cite{Klebanov:1998hh,Klebanov:2000nc,Klebanov:2000hb}. 

The orientifold projection we wish to consider corresponds to adding an O7-plane embedded via the equation $z_{4} = 0$. This reduces the $SO(4)$ isometry of \cref{eq: conifold} to $SO(3)$. The fixed locus under the orientifold spans $\mathbb{R}^4 \times M_4$, where $M_4 \subset \mathcal{C}$ is a complex two-dimensional hypersurface within the conifold. This embedding preserves the same supercharges as the D3-branes \cite{Kuperstein:2004hy}, therefore the effective four-dimensional gauge theory obtained on a stack of (regular and fractional) D3-branes in the presence of the orientifold is again a ${\cal N}=1$ quiver gauge theory. 
The field theory has gauge group, matter content and superpotential (see for instance \cite{Naculich:2002pq})\footnote{An efficient way to study orientifold projections at (toric) Calabi-Yau singularities is by using the dimer technology. See, e.g., \cite{Franco:2007ii,Argurio:2019eqb,Argurio:2020dkg, Argurio:2020npm,Argurio:2020dko,Garcia-Valdecasas:2021znu,Argurio:2022vfq} for applications of dimer technology to orientifolds.}
\begin{align} 
\begin{split}
    &\text{Gauge group:}\qquad  \, USp(N^{(1)}) \times USp(N^{(2)}) \\ 
    &\text{Matter fields:}  \qquad (X_i)^{a}_{\dot{a}} \in (\fund,\fund,\fund) \\ 
    &\text{Superpotential:} \quad \, \, \mathcal{W} = h J_{ab}J_{cd} J^{\db\dc}J^{\dd\da}\left[ (X_1)^a_\da (X_1)^b_\db (X_2)^c_\dc (X_2)^d_\dd - (X_1)^a_\da (X_2)^b_\db (X_2)^c_\dc (X_1)^d_\dd \right] \ ,\label{W tree level 1}
    \end{split}
\end{align}
where we have explicitly written the $USp(N^{(1)}) \times USp(N^{(2)})$ indices $(a,\dot{a})$, the $SU(2)_F$ flavor index $i=1,2$ and $J$ is the invariant tensor of the unitary symplectic group. The $SU(2)_F$  flavor symmetry matches the $SU(2) \simeq SO(3)$ isometry of the orientifolded conifold. We use the convention in which $N^{(1)}, N^{(2)}$ are always even. In this way, they correspond  to the number of D-branes in the parent theory.  The representation vector $(\fund,\fund,\fund)$ denotes the representation under $USp(N^{(1)}) \times USp(N^{(2)}) \times SU(2)_F$. The superpotential can be easily seen to preserve the $SU(2)_F$ flavor symmetry.
Finally, there is also a would-be $U(1)$ R-symmetry broken at the quantum level to
\begin{equation}
\label{eq: ft nonanR12}
    U(1)_R \, \rightarrow \, 
    \mathbb{Z}_{gcd(N^{(1)}-N^{(2)}+2,\,N^{(1)}-N^{(2)}-2)} \ .
\end{equation}
Local gauge anomalies are trivially vanishing. Also global gauge anomalies of the Witten type \cite{Witten:1982fp} vanish because the number of flavors is always even. Thus, the quiver theory is free of (gauge) anomalies for any $N^{(1)}, N^{(2)}$, which we are therefore free to choose.

In this work we mostly focus on dynamical aspects and therefore we do not distinguish between different global forms of the gauge group. All global forms are presumably realized as different boundary conditions for gauge fields in holography (see e.g.~\cite{Bergman:2022otk,Bergman:2024its} for different holographic configurations involving symplectic gauge groups). We leave a detailed study for future work.

In the remainder of this section we will discuss the dynamics of the orientifold gauge theory. As any generic ${\cal N}=1$ quiver gauge theory on a stack of regular and fractional D3-branes at a Calabi-Yau singularity, the RG-flow is conveniently described as a Seiberg duality cascade. In what follows we first discuss the RG-flow and then focus on the two regimes of  more relevant interest, namely the deep UV and the deep IR.

\subsection{The duality cascade}

In what follows we will set, for later convenience, $N^{(1)} = N + M-2$ and $ N^{(2)} = N$ so that the gauge group is $USp(N+M-2)_1 \times USp(N)_2$. As we will now show, this theory, much like the KS model \cite{Klebanov:1998hh,Klebanov:2000nc,Klebanov:2000hb}, has a class of RG flows which are well described by a cascade of Seiberg dualities. In what follows, we want to highlight the main differences with respect to the KS theory. 

We start by computing the $\beta$-functions of the gauge couplings $\frac{8\pi^2}{g_i^2}$ and of the dimensionless superpotential coupling $\eta$, which we define from the dimensionful coupling $h$ as $\eta \equiv \mu h$, with $\mu$ the RG scale. 
Due to the symmetries, all the matter fields $(X_i)^{a}_{\dot a}$ have the same anomalous dimension, which we take to be $\gamma_0=-\frac{1}{2}+\delta_0$. 
Note that $\gamma_0=-\frac{1}{2}$ is the value for the conifold SCFT \cite{Klebanov:1998hh}. The $\beta$-functions are
\begin{equation}\label{eq: beta functions}
\begin{split}
    \beta_1&=\frac{3}{2}(N+M-2+2)-\frac{1}{2}(1-\gamma)(2N)=\frac{3}{2}M +\delta_0 N \\
    \beta_2&=\frac{3}{2}(N+2)-\frac{1}{2}(1-\gamma)(2(N+M-2))=-\frac{3}{2}(M-4)+\delta_0 (N+M-2) \\
    \beta_{\eta}&=(1+2\gamma)\eta=2\delta_0 \eta \ ,
\end{split}
\end{equation}
where $\beta_i\equiv\beta_{8\pi^2/g_i^2}$, with $i=1,2$. 

For any $M > 4$, this theory behaves exactly as the KS theory: there is no conformal manifold, but there are  
fixed points in which the superpotential and one of the two gauge couplings are turned off. Close to the 
fixed point $(g_1=g_1^*, g_2=0, \eta=0)$ the superpotential is relevant (having $\beta_1=0$ implies that $\delta_0$ is negative and so is $\beta_\eta = \mu \, d\eta/d\mu$) and draws the theory toward another fixed point, which can be better  described in a Seiberg dual frame. Being the second gauge group weakly coupled, we can regard it as a flavor group and consider an $USp(N+M-2)_1$ gauge theory with $2N$ flavors and a quartic superpotential preserving a $SU(2)_F \times USp(N)$ flavor symmetry.\footnote{This is true as long as $N_f > N_c - 4$. The dictionary maps SQCD with gauge algebra $USp(N_c)$ and $N_f$ fundamental flavors to $USp(N_f-N_c-4)$ with $N_f$ dual quarks and mesons $M$ in the antisymmetric representation of the flavor symmetry \cite{Intriligator:1995ne}.} The Seiberg dual description is
\begin{equation}
	USp(N-M-2)_1 \times USp(N)_2 \quad \text{w/} \quad  (\widetilde{X}_i)^{a}_{\dot{a}}  , \quad M_{ij,\dot{a}\dot{b}}  \ ,\label{Eq:SeibergDual}
\end{equation}
where the indices follow the same conventions as in eq.~\eqref{W tree level 1}. The superpotential is
\begin{equation}
	\mathcal{W} = \frac12\hat{h} J^{\db\dc}J^{\dd\da} \epsilon^{ik}\epsilon^{jl}M_{ij,\da\db}M_{kl,\dc\dd}
      + y M_{ij,\dot{a}\dot{b}} J^{\dot{a}\dot{c}} J^{\dot{b}\dot{d}} J_{a b} \epsilon^{i k} \epsilon^{j l} (\widetilde{X}_k)^{a}_{\dot{c}} (\widetilde{X}_l)^{b}_{\dot{d}}  \ ,
\end{equation}
where now $\hat \eta = \mu \hat h$. The first term gives the mesons a mass and we may integrate them out. The resulting superpotential is 
\begin{equation}
	\mathcal{W} = 
    -\frac{y^2}{2\hat{h}}J_{ab}J_{cd} J^{\db\dc}J^{\dd\da} \epsilon^{i k} \epsilon^{j l}(\widetilde{X}_i)^a_\da (\widetilde{X}_j)^b_\db (\widetilde{X}_k)^c_\dc (\widetilde{X}_l)^d_\dd \ ,
    \end{equation}
which has the same form of the one in the original theory, if one sets 
\begin{equation}\label{eq: cascade_shifts}\begin{split}
	&N^\prime = N-M-2 \\
	&M^\prime = M+4 \ ,
\end{split}\end{equation}
with the roles of the two gauge groups now inverted: $USp(N')_1\times USp(N'+M'-2)_2$. In this dual description, the theory now sits close to the fixed point $(g_1=g_1^*, g_2=0, \hat{\eta}=0)$. 
Here, however, the superpotential is irrelevant, while $g_2$ is relevant. The RG flow brings the theory close to the other fixed point $(g_1=0, g_2=g_2^*, \hat{\eta}=0)$, where the cascade repeats exactly as before. The story goes on and the cascade continues toward the IR. To summarize, the duality cascade from the UV to the IR goes as 
\begin{eqnarray}
\label{eq: cascade steps}
&\cdots \nn \\
&\downarrow \nn \\ 
& USp(N+M-2)_1 \times USp(N+2M-8)_2  & \Delta (\text{rank}) = M-6 \nn \\ 
&\downarrow \nn \\  &USp(N+M-2)_1 \times USp(N)_2 & \Delta (\text{rank}) = M-2  \\ 
&\downarrow \nn \\ 
& USp(N-M-2)_1 \times USp(N)_2 & \Delta (\text{rank}) = M+2 \nn \\ 
&\downarrow \nn \\ 
&\dots \nn 
\end{eqnarray}
A peculiarity of this cascade is that the difference between the gauge group ranks at each duality step,  $\Delta(\text{rank})\equiv |N^{(1)}-N^{(2)}|$, is not constant, rather it increases toward the IR. A consequence of this behavior is that this difference goes to 0 toward the UV. As observed in similar cases \cite{Argurio:2017upa}, this implies that the cascade is finite, {\it i.e.} it ends after a finite number of steps. This is in striking contrast with the $SU(N+M) \times SU(N)$ KS cascade \cite{Klebanov:2000hb}, where the difference in ranks is constant and equal to $M$ all along the flow, yielding an infinite cascade toward the UV. In particular, as we will discuss in detail in the following, in the present case the RG flow stems from a one-dimensional conformal manifold in the UV and ends with confinement in the IR, and this happens in a finite RG-time.

We end this section by considering the symmetries along the cascade. The flavor $SU(2)_F$ is obviously preserved. The non-anomalous R-symmetry \eqref{eq: ft nonanR12} for $N^{(1)}=N+M-2$, $N^{(2)}=N$ is,
\begin{equation}
\label{eq: ft nonanR}
    \mathbb{Z}_{gcd(M,M-4)}=\mathbb{Z}_{gcd(M,4)}\ ,
\end{equation}
which corresponds to $\mathbb{Z}_{4}$ if $M = 0 \text{ mod } 4$ and $\mathbb{Z}_{2}$ if $M = 2  \text{ mod } 4$. Along the cascade, $M\to M+4 k$, ensuring that the R-symmetry is preserved along the RG flow. 

\subsection{Field theory in the far UV}
\label{sec: conformal manifold}

\begin{figure}[t]
    \centering
    \begin{subfigure}{0.40\textwidth}
        \centering
        \begin{tikzpicture}[scale=0.9,baseline={(0,0)}]
            \path[use as bounding box] (0,-0.5) rectangle (6,6);
            \draw[thick] (0,0) -- (5,0);
            \draw[-{Latex[length=2mm]}, thick] (5,0) -- +(0.1,0);
            \draw[thick] (0,0) -- (0,5);
            \draw[-{Latex[length=2mm]}, thick] (0,5) -- +(0,0.1);
            \node at (5.2,-0.25) {$g_1$};
            \node at (-0.3,5.1) {$g_2$};
            \node at (3,-0.33) {$g_1^*$};
            \node at (-0.33,3) {$g_2^*$};
            \fill (3,0) circle (3pt);
            \fill (0,3) circle (3pt);
            \draw[line width= 1.6pt] (3,0) arc[start angle=0, end angle=90, radius=3cm];
        \end{tikzpicture}
        \caption{$M_{\text{UV}}=2$: the one-dimensional conformal manifold in the $\eta=0$ plane.}
  	    \label{fig: conformal manifold 1}
    \end{subfigure} 
    \hspace{0.7cm}
    \begin{subfigure}{0.40\textwidth}
    \centering
        \begin{tikzpicture}[scale=0.9,baseline={(0,0)}]
            \path[use as bounding box] (0,-0.5) rectangle (6,6);
            \draw[thick, postaction={decorate},
            decoration={markings,
            mark=at position 0.25 with {\arrow{Latex[length=2.7mm]}},
            mark=at position 0.5 with {\arrow{Latex[length=2.7mm]}},
            mark=at position 0.75 with {\arrow{Latex[length=2.7mm]}}
            }
            ] (0,0) -- (3,0);
            \draw[thick, postaction={decorate},
            decoration={markings,
            mark=at position 0.25 with {\arrow{Latex[length=2.7mm]}},
            mark=at position 0.5 with {\arrow{Latex[length=2.7mm]}},
            mark=at position 0.75 with {\arrow{Latex[length=2.7mm]}}
            }
            ] (5,0) -- (3,0);
            \draw[thick] (5,0) -- (6,0);
            \draw[-{Latex[length=2mm]}, thick] (6,0) -- +(0.1,0);
            \draw[thick, postaction={decorate},
            decoration={markings,
            mark=at position 0.25 with {\arrow{Latex[length=2.7mm]}},
            mark=at position 0.5 with {\arrow{Latex[length=2.7mm]}},
            mark=at position 0.75 with {\arrow{Latex[length=2.7mm]}}
            }
            ] (0,5) -- (0,0);
            \draw[-{Latex[length=2mm]}, thick] (0,5) -- +(0,0.1);
            \node at (6.2,-0.25) {$g_1$};
            \node at (-0.25,5.1) {$\eta$};
            \node at (3,-0.33) {$g_1^*$};
            \fill (3,0) circle (3pt);
            \draw[thick, postaction={decorate},
            decoration={markings,
            mark=at position 0.2 with {\arrow{Latex[length=2.7mm]}},
            mark=at position 0.4 with {\arrow{Latex[length=2.7mm]}},
            mark=at position 0.6 with {\arrow{Latex[length=2.7mm]}},
            mark=at position 0.8 with {\arrow{Latex[length=2.7mm]}}
            }
            ] (3, 0) -- (4,5);
            \draw[thick, postaction={decorate},
            decoration={markings,
            mark=at position 0.2 with {\arrow{Latex[length=2.7mm]}},
            mark=at position 0.4 with {\arrow{Latex[length=2.7mm]}},
            mark=at position 0.6 with {\arrow{Latex[length=2.7mm]}},
            mark=at position 0.8 with {\arrow{Latex[length=2.7mm]}}
            }
            ] (0.1, 5) .. controls (0,-1) and (3,0) .. (3.9,5);
            \draw[thick, postaction={decorate},
            decoration={markings,
            mark=at position 0.2 with {\arrow{Latex[length=2.7mm]}},
            mark=at position 0.4 with {\arrow{Latex[length=2.7mm]}},
            mark=at position 0.6 with {\arrow{Latex[length=2.7mm]}},
            mark=at position 0.8 with {\arrow{Latex[length=2.7mm]}}
            }
            ] (0.3, 5) .. controls (0.5,1) and (2.5,0) .. (3.7,5);
            \draw[thick, postaction={decorate},
            decoration={markings,
            mark=at position 0.2 with {\arrow{Latex[length=2.7mm]}},
            mark=at position 0.4 with {\arrow{Latex[length=2.7mm]}},
            mark=at position 0.6 with {\arrow{Latex[length=2.7mm]}},
            mark=at position 0.8 with {\arrow{Latex[length=2.7mm]}}
            }
            ] (6, 2) .. controls (4,3) and (4.1,4) .. (4.2,5);
            \draw[thick, postaction={decorate},
            decoration={markings,
            mark=at position 0.2 with {\arrow{Latex[length=2.7mm]}},
            mark=at position 0.4 with {\arrow{Latex[length=2.7mm]}},
            mark=at position 0.6 with {\arrow{Latex[length=2.7mm]}},
            mark=at position 0.8 with {\arrow{Latex[length=2.7mm]}}
            }
            ] (6, 1) .. controls (3.7,1.5) and (4,4) .. (4.1,5);
        \end{tikzpicture}
        \caption{RG flow close to a fixed point where $\eta$ is a relevant deformation.}
        \label{fig: conformal manifold 2}
    \end{subfigure} 
    
    \vspace{0.5cm}
    \begin{subfigure}{0.40\textwidth}
        \centering
        \begin{tikzpicture}[scale=0.9]
            \path[use as bounding box] (0,-0.8) rectangle (6,5);
            \draw[thick] (0,0) -- (5,0);
            \draw[thick] (0,0) -- (0,5);
            \draw[-{Latex[length=2mm]}, thick] (5,0) -- +(0.1,0);
            \draw[-{Latex[length=2mm]}, thick] (0,5) -- +(0,0.1);
            \node at (5,-0.33) {$g_1$};
            \node at (-0.33,5) {$g_2$};
            \node at (3,-0.33) {$g_1^*$};
            \node at (-0.33,3) {$g_2^*$};
            \fill (3,0) circle (3pt);
            \fill (0,3) circle (3pt);
            \draw[thick, postaction={decorate},
            decoration={markings,
            mark=at position 0.2 with {\arrow{Latex[length=2.7mm]}},
            mark=at position 0.4 with {\arrow{Latex[length=2.7mm]}},
            mark=at position 0.6 with {\arrow{Latex[length=2.7mm]}},
            mark=at position 0.8 with {\arrow{Latex[length=2.7mm]}}
            }
            ] (0,3) arc[start angle=90, end angle=0, radius=3cm];
        \end{tikzpicture}
        \caption{For $M_{\text{UV}}=4$, $g_2$ is an irrelevant deformation of the fixed point $g_1=g_1^*, g_2=0, \eta=0$.\\}
        \label{fig: conformal manifold 3}
    \end{subfigure} 
    \hspace{0.7cm}
    \begin{subfigure}{0.40\textwidth}
        \centering
        \begin{tikzpicture}[scale=0.9]
            \path[use as bounding box] (0,-0.8) rectangle (6,6);
            \draw[thick] (0,0) -- (5,0);
            \draw[-{Latex[length=2mm]}, thick] (5,0) -- +(0.1,0);
            \draw[thick] 
            (0,5) -- (0,0);
            \draw[-{Latex[length=2mm]}, thick] (0,5) -- +(0,0.1);
            \node at (5,-0.33) {$g_2$};
            \node at (-0.25,5) {$\eta$};
            \node at (3,-0.33) {$g_2^*$};
            \fill (3,0) circle (3pt);
            \node at (5.3,2.3) {{\color{red}$S$}};
            \begin{scope}
                \coordinate (A) at (3,0);
                \coordinate (B) at (5,4.3);
                \coordinate (C1) at (3.2,1);
                \coordinate (C2) at (4,3.8);
                \draw[line width=1.6pt] (A) .. controls (C1) and (C2) .. (B);
            \end{scope}
            \begin{scope}
                \coordinate (A) at (3.6,0.7);
                \coordinate (B) at (4.9,3.8);
                \coordinate (C1) at (5,1);
                \coordinate (C2) at (5.5,3.3);
                \draw[thick, red] (A) .. controls (C1) and (C2) .. (B);
                \draw[-{Latex[length=2.5mm]}, thick, red] (A) -- +(-0.35,-0.05);
                \draw[-{Latex[length=2.5mm]}, thick, red] (B) -- +(-0.1,+0.15);
            \end{scope}
        \end{tikzpicture}
        \caption{$M_{\text{UV}}=4$: the one-dimensional conformal manifold in the $g_1=0$ plane. Seiberg duality acts as an S-duality on $\eta$.}
        \label{fig: conformal manifold 4}
    \end{subfigure} 
    \caption{Analyzing the RG flow around the UV conformal manifold of the orientifold gauge theory.}
    \label{fig:conformal manifold}
\end{figure}
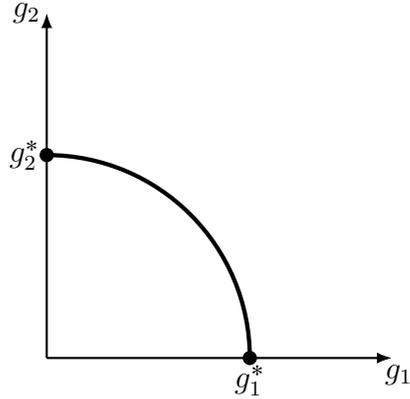
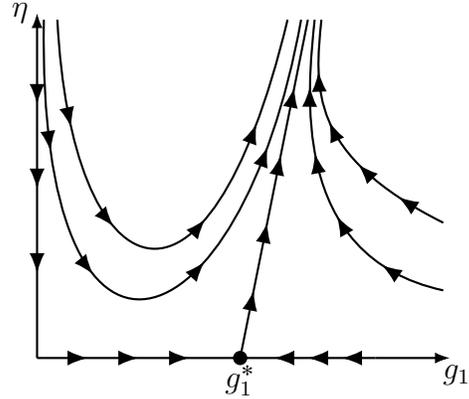
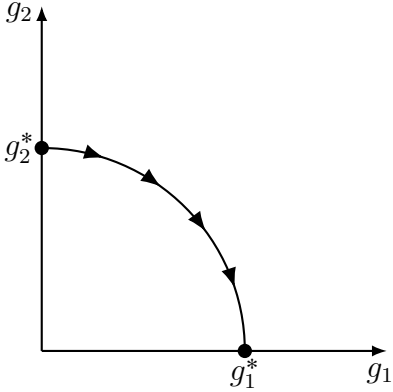
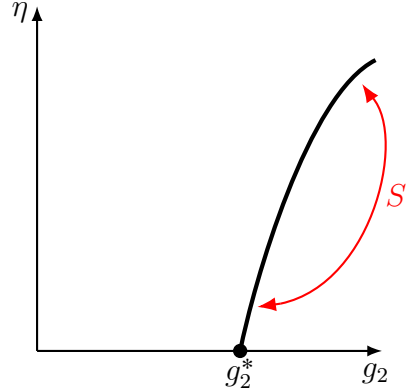

As we have seen, $\Delta(\text{rank})$ increases by $4$ at each cascade step toward the IR which is equivalent to saying that it decreases by the same amount when moving toward high energies. This implies that, in the far UV, the cascading flow will be described by one of two possible gauge theories: $USp(N_{\text{UV}})_1\times USp(N_{\text{UV}})_2$ or $USp(N_{\text{UV}}+2)_1\times USp(N_{\text{UV}})_2$ (recall that $N$ and $M$ are both even). To understand why in both cases the cascade stops after a finite number of steps, we need to evaluate the $\beta$-functions in \eqref{eq: beta functions} for $M=M_{\text{UV}}=2$ and $M=M_{\text{UV}}=4$, respectively. Let us discuss the two cases separately.
\begin{itemize}
    \item $M_{\text{UV}} = 2$. In this case we have
    \begin{equation}\label{eq: beta functions M=0}
    \begin{split}
        \beta_1&=3+\delta_0 N_{\text{UV}}\ , \\
        \beta_2&=3+\delta_0 N_{\text{UV}}\ , \\
        \beta_{\eta}&=2\delta_0\eta\ .
    \end{split}
    \end{equation}
    Naively, one might want to seek a UV free fixed point. However, this is in contrast with the need of a large negative anomalous dimension $\gamma_0 \lesssim -1/2$ for small (gauge and quartic) couplings. Instead, observing that the $\beta$ functions of the two gauge couplings are equal, we conclude that there exists a one-dimensional conformal manifold defined by the equations
    \begin{equation}
    \label{an1}
        \eta=0\ , \, \quad \quad \gamma_0(g_1,g_2,\eta=0)=-\frac{1}{2}-\frac{3}{N_{\text{UV}}}\ .
    \end{equation}
    This is schematically depicted in figure \ref{fig: conformal manifold 1}. The superpotential coupling is relevant all along the conformal manifold, and, in particular, it is relevant close to the two fixed points $(g_1=g_1^*, g_2=0, \eta=0), (g_1=0, g_2=g_2^*, \eta=0)$, triggering the cascade. The situation is depicted in figure \ref{fig: conformal manifold 2} for the first of these two fixed points. Notice, further, that since $\eta$ is relevant close to both fixed points, the cascade cannot proceed toward the UV. 

    \item $M_{\text{UV}}=4$. In this case we have
    \begin{equation}\label{eq: beta functions M=2}
    \begin{split}
        \beta_1&=6+\delta_0 N_{\text{UV}}\ , \\
        \beta_2&=\delta_0 (N_{\text{UV}}+2)\ , \\
        \beta_{\eta}&=2\delta_0 \eta\ .
    \end{split}
    \end{equation}
Here we still seemingly have two fixed points. The first one, $(g_1=g_1^*, g_2=0, \eta=0)$ is actually very similar to the ones at lower steps of the cascade: $\eta$ is relevant and triggers the cascading flow toward the IR (see figure \ref{fig: conformal manifold 2}), while $g_2$ is irrelevant, so, following this deformation toward the UV, one ends up on another fixed point $(g_1=0, g_2=g_2^*, \eta=0)$ (figure \ref{fig: conformal manifold 3}). From \eqref{eq: beta functions M=2}, it is apparent that this second fixed point is part of a one dimensional conformal manifold, parameterized by 
\begin{equation}
\label{an2}
g_1=0\ , \quad \quad \gamma(g_1=0,g_2,\eta)=-\frac{1}{2}\ .
\end{equation}
This is depicted in figure \ref{fig: conformal manifold 4}. On this conformal manifold $g_1$ is relevant and hence triggers the cascading flow toward the IR. Conversely, also in this case, the cascade cannot proceed toward the UV.

As a side remark, we note that on such a conformal manifold an exact S-duality is expected to act, mapping the region of small $\eta$ to the region of large $\eta$. This is the same situation one has in $\mathcal{N}=1$ $SU(N_c)$ SQCD with $N_f=2N_c$ flavors and a quartic superpotential, which admits an exactly marginal deformation and is self-dual under Seiberg duality \cite{Strassler:2005qs}. This is depicted in figure \ref{fig: conformal manifold 4}. Indeed, if we were to perform a Seiberg duality close to this fixed point taking $USp(N_{\text{UV}}+2)_1$, which is weakly coupled, to be a flavor symmetry, we would get
\begin{equation}
    USp(N_{\text{UV}})_2\to USp(2(N_{\text{UV}}+2)-N_{\text{UV}}-4)_2=USp(N_{\text{UV}})_2 \ .
\end{equation}
So the two end points of the conformal manifold are the same point. 
\end{itemize}

\subsection{Field theory in the deep IR}
\label{sec: field theory IR}

The moduli space of our orientifold gauge theory is very large and stratified, consisting of many branches as is typical for cascading RG-flows \cite{Dymarsky:2005xt, Argurio:2007vq}. Our aim will be to find the smallest branch of the moduli space, so that our main focus will be on VEVs of gauge invariant massless degrees of freedom at the bottom of the cascade. 

On general grounds, since there are no baryons for $USp$ gauge groups, one could expect just mesonic branches and, in turn, no choice of gauge group ranks for which a non-singular, purely geometric, dual supergravity description exists (unlike for the KS model which admits a purely geometric baryonic branch, whenever $N$ is a multiple of $M$). As we are going to show below, this expectation is naive and a purely geometric mesonic vacuum exists. 

Let us consider our quiver gauge theory with gauge group $USp(N+M-2)_1\times USp(N)_2$. At the effective level, we can consider gauge invariants of the most strongly coupled gauge group, which we take to be the one of larger rank. These are,
\begin{equation}
(M_{ij})_{\da\db} =J_{ab} (X_i)^a_\da (X_j)^b_\db\ .
\end{equation}
For the analysis which follows, we find it more convenient to use a matrix notation. We can build a $2N\times 2N$ matrix as
\begin{equation}
\cM = \begin{pmatrix}
    M_{11} & M_{12} \\ M_{21} & M_{22}
\end{pmatrix}\ ,
\end{equation}
which has the property that $\cM^T=-\cM$, implying $M_{11}^T=-M_{11}$, $M_{22}^T=-M_{22}$ and $M_{12}^T=-M_{21}$.

The strongly coupled gauge group has $N_c= N+M-2$ and $N_f= 2N$. Let us first consider the case $N\geq M+4$, which means $N_f \geq N_c +6$. Assuming that all mesons have large VEVs, in matrix notation the full effective superpotential on this branch of moduli space reads \cite{Intriligator:1995ne}
\begin{equation}\label{weff}
\cW= h \Tr (M_{11} J M_{22} J - M_{12} J M_{21} J) + c\Lambda^3  \left( \frac{\det \cM}{\Lambda^{4N}}\right)^\frac{1}{N-M}\ ,
\end{equation}
for some constant $c$ that is not important in the following.  
Notice that since $\cM$ is antisymmetric, we can define its Pfaffian, and we normalize it such that $\det \cM= (\Pf \cM)^2$. Hence we can write any correction involving the Pfaffian as a power of the determinant.

The F-terms of this effective theory are now 
\begin{equation}
hJ\begin{pmatrix}
    M_{22} & -M_{12} \\ -M_{21} & M_{11} 
\end{pmatrix}J + c'\Lambda^3  \left( \frac{\det \cM}{\Lambda^{4N}}\right)^\frac{1}{N-M} (\cM)^{-1}=0\ .
\end{equation}
Multiplying by $\cM$ either from the left or the right the above equations (and if needed by $J$), one gets that
\begin{equation}
\label{commutationsMJ}
M_{ij}J M_{kl}J=M_{kl}J M_{ij}J,
\end{equation}
for all pairs of indices, and that 
\begin{equation}
M_{11}JM_{22}J-M_{12}J M_{21}J = -h^{-1}c'\Lambda^3  \left( \frac{\det \cM}{\Lambda^{4N}}\right)^\frac{1}{N-M}\ .
\end{equation}
The relations \eqref{commutationsMJ} imply that one can simultaneously diagonalize all the $N\times N$ matrices $M_{ij}J$. Notice, however, that $M_{11}$ and $M_{22}$ are antisymmetric, which translates to $(M_{11}J)^T=-J(M_{11}J)J$ and similarly for $M_{22}$. Now, take $J$ to be block-diagonal with $2\times 2$ blocks given by $\begin{pmatrix}
    0 & 1 \\ -1 & 0
\end{pmatrix}$. Hence eigenvalues of $M_{11}J$ and $M_{22}J$ should come in pairs, with each eigenvalue appearing twice in each block. As for $M_{12}$ and $M_{21}$, their relation $M_{12}^T=-M_{21}$ implies $(M_{12}J)^T=-J(M_{21}J)J$, so that the eigenvalues of the two matrices are related as we now show. Decomposing $M_{12}$ and $M_{21}$ in the same blocks as the other matrices, we get that 
\begin{equation}
\mathrm{if} \quad M_{12}J= \begin{pmatrix}
    m_{12} & 0 \\ 0 & m_{21}
\end{pmatrix}\ , \quad \mathrm{then} \quad
M_{21}J= \begin{pmatrix}
    m_{21} & 0 \\ 0 & m_{12}
\end{pmatrix}\ .
\end{equation}
The remaining equations become, block by block
\begin{equation}
\label{block1212}
m_{11}^{(A)}m_{22}^{(A)}-m_{12}^{(A)}m_{21}^{(A)}= -h^{-1}c'\Lambda^3  \left( \frac{\det \cM}{\Lambda^{4N}}\right)^\frac{1}{N-M}\ ,
\end{equation}
with $A=1, \dots N/2$.
It is easy to convince oneself that, using the same block diagonal form, one has
\begin{equation}
\det \cM \propto \prod_{A} (m_{11}^{(A)}m_{22}^{(A)}-m_{12}^{(A)}m_{21}^{(A)})^2\ . 
\end{equation}
Taking the product over $A$ of eq.~\eqref{block1212}, one ultimately determines the value of $\det \cM$ as a function of $\Lambda$ and $h$. Eventually, one gets the deformed conifold equation $N/2$ times.\footnote{This becomes more transparent when using an alternative parametrization of the conifold, compared to eq.~\eqref{eq: conifold}, namely $xy-wz=0$, and $xy-wz=\epsilon^2$ for the deformed conifold. The two parametrizations are related by a simple change of coordinates.} In the dual type IIB picture this corresponds to $N/2$ mobile D3-branes probing the deformed 
conifold.\footnote{ As in \cite{Dymarsky:2005xt}, the moduli space consists of several copies of the same (symmetric product) manifold labeled by the phases of the gaugino condensate of a $USp$ subgroup of the gauge group. This will not be of importance to us, as the supergravity solution will describe only one of those copies.}

This is just one branch of the moduli space, the largest one for a given $N$. The other, smaller, branches are obtained when one does not assume large VEVs for $\cM$. In that case, the appropriate description when the $USp(N+M-2)$ gauge group goes at strong coupling is to perform a Seiberg/Intriligator-Pouliot duality \cite{Intriligator:1995ne} that effectively reduces its rank, and introduces dual bifundamental fields $\widetilde X_i$, so that the story repeats itself. Therefore, as in the KS theory, there is a mesonic branch at each cascade step. 

This occurs until the possibility of performing a further duality no longer exists, {\it i.e.} when the ranks are such that $N_f\leq N_c+4$, that is $N\leq M+2$. Consider first $N<M$. In this case we still have an effective superpotential that reads like \eqref{weff}, though now the power of $\det \cM$ is negative. Nevertheless, the analysis goes through exactly as before and again one gets as moduli space $N/2$ copies of the deformed conifold.

When $N=M$, the quantum corrected superpotential is actually a constraint
\begin{equation}
\cW= h \Tr (M_{11} J M_{22} J - M_{12} J M_{21} J) + L (\Pf \cM-\Lambda^{2N})\ ,
\end{equation}
where we have preferred now to use $\Pf\cM$ instead of $\det\cM$. Most of the reasoning goes on as before, until we get the relation for each set of eigenvalues
\begin{equation}
m_{11}^{(A)}m_{22}^{(A)}-m_{12}^{(A)}m_{21}^{(A)}= -\frac12 h^{-1} L \Pf \cM =  -\frac12 h^{-1} L \Lambda^{2N}=\Lambda^4\ ,
\end{equation}
where in the second equality we have implemented the constraint coming from extremizing with respect to $L$, while in the third equality we have used the fact that all equations are the same. In particular, we get $L=-2h\Lambda^{4-2N}$. Note that the constraint tells us that also at this (last) step of the cascade, there is a mesonic branch consisting of $N/2=M/2$ copies of the deformed conifold.

We are left with what turns out to be the most interesting case, namely $N=M+2$. In this case, for any value of $\cM$ the quantum corrected superpotential is given by 
\begin{equation}
\cW= h \Tr (M_{11} J M_{22} J - M_{12} J M_{21} J) + c\Lambda^3  \left( \frac{\det \cM}{\Lambda^{4N}}\right)^\frac{1}{2}\ ,
\end{equation}
which can be rewritten as 
\begin{equation}
\cW= h \Tr (M_{11} J M_{22} J - M_{12} J M_{21} J) + c \frac{\Pf \cM}{\Lambda^{2N-3}}\ .
\end{equation}
This looks deceptively similar to the previous case, since the correction is linear in $\Pf\cM$. Indeed, we still have a mesonic branch in which $\Pf \cM$ is fixed in terms of $h$ and $\Lambda$, and that corresponds once more to $N/2=(M+2)/2$ copies of the deformed conifold. However, the F-term equations derived from the above superpotential allow also for solutions where $\Pf\cM=0$ (which is forbidden by the constraint in the previous case). 
In fact, the condition $\Pf\cM=0$ implies $\cM=0$.
The reasoning goes as follows. First of all, assume for simplicity that the matrices $M_{ij}J$ have all been diagonalized. Then the effective superpotential reads
\begin{equation}
    \cW= h\sum_A (m_{11}^{(A)}m_{22}^{(A)}-m_{12}^{(A)}m_{21}^{(A)}) +\frac{c}{\Lambda^{2N-3}}\prod_A (m_{11}^{(A)}m_{22}^{(A)}-m_{12}^{(A)}m_{21}^{(A)})\ .
\end{equation}
The F-term equations are then
\begin{equation}
    h\ m_{ij}^{(B)} = - \frac{c}{\Lambda^{2N-3}}\prod_{A\neq B} (m_{11}^{(A)}m_{22}^{(A)}-m_{12}^{(A)}m_{21}^{(A)})\ m_{ij}^{(B)}\ .
\end{equation}
If $\Pf\cM\neq 0$, we obtain the solution discussed above. If on the other hand $\Pf\cM=0$, then it means that there is at least one value of the index $A$ such that $m_{11}^{(A)}m_{22}^{(A)}-m_{12}^{(A)}m_{21}^{(A)}=0$. The equation above then implies that $m_{ij}^{(B)}=0$ for any $B\neq A$. This in turn implies that also $m_{ij}^{(A)}=0$, hence concluding that $\cM=0$.
This is clearly a disconnected point of the vacuum manifold, actually the smallest (zero-dimensional) branch that one can achieve. Note that this point-like branch is the only one where the $SU(2)$ symmetry is preserved, since in any other branch the mesons, which transform under this symmetry, necessarily have non-zero VEVs. 

In conclusion, a generic cascade that ends up with $USp(N+M-2) \times USp(N)$ gauge groups has a mesonic branch which corresponds to $N/2$ D3-branes probing the deformed conifold. Given that a large number of mobile D3-branes should backreact and leave a completely smooth geometry, the gravity dual of such vacua should be similar to the ones discussed in \cite{Krishnan:2008gx} for the KS theory, with an $AdS_5\times S^5$ throat opening-up near the bunch of mobile D3s. An exception to this situation is the vacuum at the origin of the moduli space which exists when $N=M+2$. In this vacuum all mesons are massive and can be integrated out, and one is left with pure $USp(M+2)$ SYM, which confines. This is the only vacuum whose dual background is expected to be purely geometrical, similarly to the baryonic branch of the KS theory. In the following, we will argue that the supergravity solution we find describes exactly such vacuum.

\subsection{The entire RG flow: wrap-up}
\label{sec: full RG}

As we have shown, the RG flow can be described by a duality cascade that interpolates from a UV fixed point—more precisely, a one-dimensional conformal manifold—down to the IR, where the gauge theory vacua lie on a mesonic branch. The cascade is finite in the sense that, starting from a given effective gauge theory in the IR and climbing the cascade toward the UV (or vice versa), after a finite number of duality steps the cascade stops. With a slight abuse of language, we say that the cascade ends after a finite RG time. This can be made concrete by showing that there is a precise relation between the gauge group ranks in the UV and in the IR.

Let us suppose that the ranks of the two gauge groups at the bottom of the cascade are $N^{(1)}_0=N_0+M_0-2$ and $N^{(2)}_0=N_0$. From the analysis of section \ref{sec: conformal manifold}, we know that, following the cascade toward the UV, the difference of the ranks of the two gauge groups decreases by $4$ at each cascade step: the cascade stops when this difference is either $0$ or $2$. Let us indicate with $N^{(1)}_n=N_n+M_n-2$ and $N^{(2)}_n=N_n$ the ranks of the gauge groups after $n$ steps toward the UV. Inverting \eqref{eq: cascade_shifts}, one can readily  express $N_n$ and $M_n$ in terms of $N_0$ and $M_0$, obtaining the following result
\begin{equation}\label{eq: shift at step k}\begin{split}
M_{n}&=M_{n-1}-4=...=M_0-4n  \\ 
N_{n}&=N_{n-1}+M_{n-1}-2=N_{n-2}+M_{n-2}+M_{n-1}-4=...=N_0+\sum_{i=0}^{n-1} M_i-2n= \\
&=N_0+ \sum_{i=0}^{n-1} (M_0-4i)-2n=N_0+n\,M_0-2n^2 +2n -2n=N_0+n\,M_0-2n^2\ . 
\end{split}\end{equation}
To determine the ranks in the UV, we need to use the first equation in \eqref{eq: shift at step k} to compute the number of cascade steps. Then, we can substitute the number of steps in the second equation to evaluate $N_{\text{UV}}$. As discussed in section \ref{sec: conformal manifold}, we have only two possible cases:

\begin{itemize}
    \item $M_{\text{UV}}=2$. From the first equation in \eqref{eq: shift at step k}, we find:
    \begin{equation}
        M_n=M_{\text{UV}}=2 \rightarrow n=\frac{M_0-2}{4}
    \end{equation}
    The gauge groups in the UV are $USp(N_{\text{UV}})_1\times USp(N_{\text{UV}})_2$, with:
    \begin{equation}
    \label{eq: N0M0_1}
        N_{\text{UV}}=N_0+\frac{M_0^2}{8}-\frac{1}{2} \ . 
    \end{equation} 
    
    \item $M_{\text{UV}}=4$. Once again, from the first equation in \eqref{eq: shift at step k}, we find:
    \begin{equation}
        M_n=M_{\text{UV}}=4 \rightarrow n=\frac{M_0-4}{4}
    \end{equation}
    The gauge groups in the UV are $USp(N_{\text{UV}}+2)_1\times USp(N_{\text{UV}})_2$, with:
    \begin{equation}
    \label{eq: N0M0_2}
        N_{\text{UV}}=N_0+\frac{M_0^2}{8}-2 \ . 
    \end{equation} 
\end{itemize}
So we conclude that the dependence of the gauge group ranks in the UV on IR data is linear in $N_0$ and quadratic in $M_0$.

\section{Supergravity solution}
\label{sec: SUGRA solution}

The $\mathcal{N}=1$ quiver gauge theory with gauge group $USp(N^{(1)})\times USp(N^{(2)})$ is obtained as the low-energy effective theory describing the dynamics of a stack of regular and  fractional $D3$-branes at the tip of the orientifolded conifold geometry. The non-orientifolded version of this theory is a quiver gauge theory which has been extensively studied in the literature, starting from \cite{Klebanov:1998hh,Klebanov:2000nc, Klebanov:2000hb}. Our proposal for the gravity dual of the orientifold quiver gauge theory is a deformation of these backgrounds, as we now explain. 

The KS solution \cite{Klebanov:2000hb} has constant $F_3$ flux, proportional to the number of fractional D3-branes, logarithmically running $F_5$ and $H_3$ fluxes, vanishing $F_1$ flux and constant dilaton. The O7-plane sources D7 charge and couples to the dilaton. In this sense, from the point of view of coupling to supergravity fields, the O7-plane behaves very similarly to a stack of $8$ D7-branes, with opposite tension and charge. The problem of adding D7 branes to the Klebanov-Strassler and Klebanov-Tseytlin backgrounds was studied in \cite{Benini:2006hh,Benini:2007gx,Benini:2007kg}. Within the gauge/gravity correspondence the introduction of the D7-branes corresponds to the introduction of flavors in the dual gauge theory \cite{Grana:2001xn,Bertolini:2001qa,Karch:2002sh}. There are several possible supersymmetric embeddings for D7-branes in the conifold geometry \cite{Ouyang:2003df, Arean:2004mm, Kuperstein:2004hy}. One of them, the so called Kuperstein embedding \cite{Kuperstein:2004hy}, coincides with the embedding of the O7-plane discussed in the previous section. 
This suggests that the solution we are after should enjoy the same supersymmetric ansatz of \cite{Benini:2007gx}, where the addition of D7-branes with Kuperstein embedding on the conifold was discussed, with the caveat that we will substitute $N_f \to -8$.

\subsection{Ansatz}
\label{sec: ansatz}

Our ansatz is identical to that of \cite{Benini:2007gx} which, for ease of comparison, we now review. The ten-dimensional spacetime coordinates are
\begin{equation}\{x^0,x^1,x^2,x^3, r, \theta_1,\varphi_1,\theta_2,\varphi_2,\psi\} \,
\end{equation}
where the four coordinates $x^\mu, \,\mu=0,1,2,3$, parameterize Minkowski spacetime and the other six coordinates, with ranges $r \geq 0$, $0 \leq \theta_i\leq \pi$, $0 \leq \varphi_i\leq 2\pi$, $0 \leq \psi\leq 4 \pi $, parameterize the conifold geometry, a real cone over $T^{1,1}$. Topologically, the base of the conifold is $S^2 \times S^3$ where $S^2$ is defined by $\theta_1=\theta_2$, $\varphi_1=2\pi - \varphi_2$ and fixed $\psi$, while $S^3$ by $\theta_2=\text{const}$, $\varphi_2=\text{const}$. The orientifold acts as a $\mathbb{Z}_2$ involution on spacetime.  For simplicity, in deriving the supergravity solution we work in the covering space and incorporate the effects of the quotient later, when performing the gauge/gravity duality checks in section \ref{sec: checks}.

The ansatz for the metric (which we take to be in the Einstein frame, as in \cite{Benini:2007gx}) is 
\begin{equation}
    \label{eq: metric Ansatz}
    \begin{split}
        ds^2=&h(r)^{-\frac{1}{2}}dx_{1,3}^2+ h(r)^\frac{1}{2}\bigg[dr^2+e^{2G_1(r)} (\sigma_1^2+\sigma_2^2)+e^{2G_2(r)}((\omega_1+g(r)\sigma_1)^2+\\&(\omega_2+g(r)\sigma_2)^2) 
        +\frac{e^{2G_3(r)}}{9}(\omega_3+\sigma_3)^2
        \bigg] \, ,
    \end{split}
\end{equation}
where $dx_{0,3}^2$ is the Minkowski metric and we have defined the following set of one forms
\begin{align}
\label{sigmaomega}
    &\sigma_1=d\theta_1 \quad , \quad  \sigma_2=\sin{\theta_1}d\varphi_1 \quad ,  \quad \sigma_3=\cos{\theta_1}d\varphi_1 \quad , \quad \omega_1=\sin{\psi}\sin{\theta_2}d\varphi_2+\cos{\psi}d\theta_2 \nonumber\\
&\omega_2=-\cos{\psi}\sin{\theta_2}d\varphi_2+\sin{\psi}d\theta_2 \quad , \quad \omega_3=d\psi+\cos{\theta_2}d\varphi_2 \ ,
\end{align}
and from now on, we set $\alpha'=g_s=1$. Notice that the presence of the O7 plane violates the Bianchi identity for the $F_1$ RR form. Following \cite{Benini:2007gx}, to simplify the computations it is convenient to smear the D7-branes (actually the O7-plane in our case) in their two transverse directions. In this way the violation of the Bianchi identity is independent of both the radial and the $\psi$ (fiber) coordinates, and it is also symmetric under the exchange between $(\theta_1,\varphi_1)$ and $(\theta_2,\varphi_2)$. This condition is required by the fact that the latter are exchanged by the geometrical action of the orientifold, and hence identified. 
We assume that the smearing does not change in a dramatic way what we want to show. The detailed matching that we will provide in the following justifies {\em a posteriori} this assumption.\footnote{The smearing can be seen as a truncation to the zero modes of the would-be localized solution. We thank Carlos Nunez for suggesting this to us.\label{footnotecarlos}} The Bianchi identity then reads 
\begin{equation}
    dF_1= -\frac{2}{\pi}(\omega_1\wedge\omega_2-\sigma_1\wedge\sigma_2) \ , 
\end{equation}
which implies
\begin{equation} \label{eq: F_1}
    F_1=-\frac{2}{\pi}(\omega_3+\sigma_3) \ . 
\end{equation}
The ansatz for the NSNS and RR 3-forms is 
\begin{equation}\begin{aligned}  \label{eq: ansatz three forms}
B_2 &= \frac{M_0}{2} \Bigl[ f\, g^1 \wedge g^2\,+\,k\, g^3 \wedge 
g^4 \Bigr]   \\
H_3 &= dB_2\,=\, \frac{M_0}{2} \, \Bigl[ dr \wedge (f' \,g^1 \wedge g^2\,+\,
k'\,g^3 \wedge g^4)\,+\,{\frac{1}{2}}(k-f)\, g^5 \wedge (g^1
\wedge g^3\,+\,g^2 \wedge g^4) \Bigr]   \\
F_3 &= \frac{M_0}{2} \Big\{ g^5\wedge \Big[ \big( F-\frac{2}{\pi}f\big)g^1\wedge g^2 + \big(1- F-\frac{2}{\pi}k\big)g^3\wedge g^4 \Big] +F' dr \wedge \big(g^1\wedge g^3 + g^2\wedge g^4   \big)\Big\} \ ,
\end{aligned} 
\end{equation}
where $M_0$ is an arbitrary constant, while $f(r)$, $k(r)$ and $F(r)$ are functions of the radial coordinate. Finally, the $g^i$'s are another convenient set of one forms, defined as
\begin{align} \label{eq: g_i definition}
    g^1&=\frac{1}{\sqrt{2}}(\omega_2-\sigma_2) & g^2&=\frac{1}{\sqrt{2}}(-\omega_1+\sigma_1) & \nonumber
    g^3&=-\frac{1}{\sqrt{2}}(\omega_2+\sigma_2) \\ g^4&=\frac{1}{\sqrt{2}}(\omega_1+\sigma_1) &
    g^5&=\omega_3+\sigma_3 \ .
\end{align}
The equations of motion must be satisfied together with the Bianchi identities, which read
\begin{equation}\label{eq: Bianchi identities}
dF_3=H_3\wedge F_1 \quad , \quad dH_3=0 \quad , \quad dF_5=H_3 \wedge F_3 \ .
\end{equation}
The equations for $F_3$ and $H_3$ are automatically satisfied by the ansatz. With the standard ansatz for $F_5$ 
\begin{equation}\label{eq: ansatz for F_5}
    F_5= (1+*) \,dh^{-1}(r)\wedge dx^0\wedge dx^1 \wedge dx^2 \wedge dx^3 \ , 
\end{equation}
its Bianchi identity gives a first order differential equation which can be integrated giving
\begin{equation}\label{eq: solution for h'}
    h'e^{2G_1+2G_2+G_3}=-\frac{3}{4}M_0^{\,2}\big[f-(f-k)F-\frac{2}{\pi}fk\big] \ . 
\end{equation}
Parameterizing $F_{5}$ in terms of its flux on a $5$-manifold at fixed radius and Minkowski coordinates, we have 
\begin{equation}
    F_{5} = \frac{\pi}{4} N_{eff}(r) g^{1} \wedge g^{2} \wedge g^{3} \wedge g^{4} \wedge g^{5} + \text{Hodge Dual} \ .
\end{equation}
We can now put this together with the ansatz \eqref{eq: ansatz for F_5} and get 
\begin{equation}
    N_{eff}(r) = - \frac{4}{3 \pi} h' e^{2G_{1} + 2G_{2} + G_{3}} \ , 
\end{equation}
which, using \eqref{eq: solution for h'}, implies
\begin{equation}\label{eq: N_eff}
    N_{eff}(r) = \frac{1}{(4 \pi^2)^2} \int_{T^{1,1}} 
    F_{5} = N_0+
    \frac{M_0^{\, 2}}{\pi} \biggl( f - (f - k) F - \frac{2}{\pi} f k \biggr) \ ,
\end{equation}
where
\begin{equation}\label{eq:volume_5}
    \int_{T^{1,1}} 
    g^1 \wedge g^2 \wedge g^3 \wedge g^4 \wedge g^5 = (4 \pi)^3 \ ,
\end{equation}
while $N_0$ is, for now, an arbitrary integration constant.\footnote{The constants $N_0$ and $M_0$ defined in eq.~\eqref{eq: ansatz three forms} will later be shown to correspond to the same quantities defined in section \ref{sec: full RG}. Therefore, to avoid clutter, we use the same notation from the onset.} Integrating our ansatz for $F_{3}$ over the $S^{3}$ given by $\theta_{2}=\text{const}$, $\varphi_{2}=\text{const}$, and using 
\begin{equation}\label{eq:volume_3}
    \int_{S_3} 
    g^5 \wedge g^1 \wedge g^2 = \int_{S_3} 
    g^5 \wedge g^3 \wedge g^4 = 8\pi^2 \ ,
\end{equation}
we get
\begin{equation}\label{eq: M_eff}
    M_{eff}(r) = \frac{1}{4 \pi^2} \int_{S^{3}} F_{3} = M_0 \biggl(1 - \frac{2}{\pi} (f + k) \biggr) \ .
\end{equation}
$N_{eff}$ and $M_{eff}$ are the so-called Maxwell charges, which are gauge invariant but generally not quantized \cite{Marolf:2000cb}.

\subsection{Solution in IR and UV regimes}
\label{sec: solution in IR/UV}

A derivation of the full solution can be found in Appendix \ref{app: SUGRA solution}, to which we refer the reader for details. In this section,  we highlight some of its key features that are essential for the gauge/gravity duality checks to be performed later, focusing on the two most relevant regimes, corresponding to the UV and the IR limits of the dual gauge theory.

Despite having a similar ansatz, the supergravity solution of \cite{Benini:2007gx} and ours have a few crucial differences which are in one-to-one correspondence with how different the orientifold gauge theory and the flavored KS gauge theory discussed in \cite{Benini:2007gx} behave in the UV and in the IR. 

The first difference is that in the flavored theory the sum of the inverse squared gauge couplings goes to zero in the UV. This gives rise to an accumulation point of Seiberg dualities, {\it i.e.} a duality wall, making the gauge theory ill-defined in the UV. The orientifold gauge theory has the opposite behavior, in that the sum of the inverse squared gauge couplings becomes small in the IR and large in the UV. However, before hitting the duality wall, which would be an IR effect in our case, the gauge theory confines and such pathological behavior is absent in the orientifold gauge theory.

From the dual supergravity perspective, this difference can be traced back to the dilaton profile, that, exactly opposite to what occurs in \cite{Benini:2007gx}, tends to grow at small values of the holographic coordinate and to decrease at large values. By the gauge/gravity dictionary, $e^{- \phi}$ is proportional to the sum of the inverse squared gauge couplings, therefore a divergence in the dilaton profile would signal a duality wall in the dual gauge theory. In our case the dilaton reads 
\begin{equation}
    e^\phi=\frac{\pi}{2}\frac{1}{\tau-\tau_0} \ ,
\end{equation}
where $\tau$ is defined from $r$ via $ d\tau=3e^{-G_3}dr$ and $\tau_0$ is an integration constant. For consistency, the holographic coordinate $\tau$ ranges in the interval $(\tau_0,+\infty)$. This implies that, as anticipated, the divergence occurs at small value of the radial coordinate. In fact, there is a second, physically relevant integration constant on which our solution depends, $\tau_c$, which corresponds to the value of the holographic coordinate $\tau$ at which the $S^2$ of the base of the internal manifold shrinks to a point and the $S^3$ remains finite. Choosing $\tau_c >  \tau_0$ yields a supergravity solution that interpolates between the UV region, $\tau \sim \infty$, and the deep IR at $\tau=\tau_c$, where the internal space terminates. This screens the would-be dilaton singularity, which would otherwise correspond to an IR duality wall not present in the gauge theory. As in the KS solution \cite{Klebanov:2000hb}, $\tau_c$ sets the scale where confinement occurs in the dual gauge theory. For ease of notation and without loss of generality, from now on we will set $\tau_c=0$ which implies that $\tau_0 < 0$. 

A second important difference between our solution and the flavored KS one is that while in both the effective number of regular D3-branes, measured by the Maxwell charge \eqref{eq: N_eff}, decreases toward the IR, in the orientifold gauge theory the fractional D3-charge (the D5-charge), measured by \eqref{eq: M_eff}, increases toward the IR, in contrast to the flavored KS model, where it decreases. This is in agreement with the field theory analysis in section \ref{sec:field theory analysis}, see in particular Eqs.~\eqref{eq: shift at step k}. We will make this matching precise later.

In the remainder of this section, we discuss the two regimes of the supergravity solution most relevant for the gauge/gravity duality checks presented in section \ref{sec: checks}, namely the large and small $\tau$ regimes. Key to this goal is the behavior of the metric of the internal space in \eqref{eq: metric Ansatz} that, in terms of  the coordinate $\tau$ reads 
\begin{equation}
    \label{eq: metric deformed conifold}
    ds^2_6=\frac{1}{2}\mu^\frac{4}{3} \tilde{\Lambda}(\tau)\left[\frac{4}{3}\frac{\tau-\tau_0}{\tilde{\Lambda}^3(\tau)}(d\tau^2+(g^5)^2)+ \cosh^2{\left(\frac{\tau}{2}\right)}((g^3)^2+(g^4)^2)+\sinh^2{\left(\frac{\tau}{2}\right)}((g^1)^2+(g^2)^2)\right] ,
    \end{equation}
where
\begin{equation}\label{eq: Lambda 1}
    \tilde{\Lambda}(\tau)=\frac{\bigg\{2\left[\sinh{(
    2 \tau)}-\tau\right](\tau-\tau_0)-\cosh \left(2\tau\right)+2\tau\tau_0+1\bigg\}^\frac{1}{3}}{\sinh{\tau}} \ .
\end{equation}
Let us note, in passing, that using eqs.~\eqref{sigmaomega} and \eqref{eq: g_i definition} one can show that the metric \eqref{eq: metric deformed conifold} enjoys $SU(2) \times SU(2)$ isometries associated to rotations in $(\theta_i,\varphi_i)$. The orientifold action identifies $(\theta_1,\varphi_1)$ with $(\theta_2,\varphi_2)$. Therefore, on the physical space $SU(2) \times SU(2)$ is broken to the diagonal $SU(2)$, nicely matching the $SU(2)_F$ flavor symmetry of the dual gauge theory, see section \ref{sec:field theory analysis}. 

\subsubsection{IR limit}
\label{sec: IR limit of SUGRA}

We want to look at the behavior of the solution in the limit $\tau\to 0$. Expanding  around $\tau \sim 0$, one can see that the metric becomes exactly the one in \cite{Klebanov:2000hb}, with a vanishing $S^2$ and a non-vanishing $S^3$ at $\tau=0$. More specifically, the metric \eqref{eq: metric deformed conifold} becomes
\begin{equation}\label{metricIR}
\begin{split}
    ds^2_6 & \underset{\tau \sim 0}{\sim} \left(\frac{\mu^4}{3}|\tau_0|\right)^\frac{1}{3}\left[\frac{1}{2}d\tau^2+\left(\frac{1}{2} (g^5)^2+(g^3)^2+(g^4)^2\right)+\frac{1}{4}\tau^2((g^1)^2+(g^2)^2)\right]= \\
    &=\left(\frac{\mu^4}{3}|\tau_0|\right)^\frac{1}{3}\left[\frac{1}{2}d\tau^2+d\Omega_3^2+\frac{1}{4}\tau^2((g^1)^2+(g^2)^2)\right] \ ,
    \end{split}
\end{equation}
where we have introduced $d\Omega_3^2$ as the volume element on the $S^3$ \cite{Minasian:1999tt}, and used the fact that for $\tau\to0$ the expression \eqref{eq: Lambda 1} reads
\begin{equation}
    \tilde{\Lambda}(\tau)= \left(\frac{8}{3}|\tau_0|\right)^\frac{1}{3}\left(1+\frac{\tau}{4|\tau_0|}+{\cal O}(\tau^2)\right)\ .
\end{equation}
The warp factor $h(\tau)$ can be obtained by plugging the solutions \eqref{eq: dilaton solution},\eqref{eq: solution deformed conifold},\eqref{eq: Lambda 1} and \eqref{eq: solution of fluxes equation} into eq.~\eqref{eq: solution for h'} and integrating. One gets
\begin{equation} \label{eq: warp factor deformed conifold}
    \begin{split}
        h(\tau)=-\frac{\pi M_0^{\, 2}}{4 \mu^\frac{8}{3}}
    &\int\limits_{\tau}^{\infty} dx \,  \frac{x\coth x-1}{(x-\tau_0)^2\sinh^2 x} \times \\
    & \times
    \frac{[-\cosh{(2x)}+4 x(x-\tau_0)+1-(x-2\tau_0)\sinh{(2x)}]}{\left[2\left(\sinh{(2x)}-x\right)(x-\tau_0)-\cosh{(2x)}+2x\tau_0+1\right]^\frac{2}{3}} \ .
    \end{split}
\end{equation}
Let us define
\begin{equation}
\label{def H(x)}
    h(\tau)=-\frac{\pi M_0^{\, 2}}{4\mu^\frac{8}{3}}
    \int\limits_{\tau}^{\infty} dx  \, H(x) \ ,
\end{equation}
and expand $H(x)$ for $x \sim 0$. We get
\begin{equation}
     H(x) \underset{x\sim 0}{\sim} -\frac{2}{3\sqrt[3]{3}}|\tau_0|^{-\frac{5}{3}} x \ .
 \end{equation}
 So, the warp factor behaves as
 \begin{equation}
     h(\tau)=-\frac{\pi M_0^{\, 2}}{4\mu^{\frac{8}{3}}} \int\limits_\tau^\infty  dx \, H(x) \underset{\tau \sim 0}{\sim} h_0+\mathcal{O}(\tau)^2 \ .
 \end{equation} 
At this scale, expanding \eqref{eq: solution of fluxes equation}, we find 
\begin{equation}
     e^{-\phi} f \sim \frac{1}{12}\tau^3, \qquad e^{-\phi} k \sim \frac{1}{3}\tau, \qquad F \sim \frac{1}{12}\tau^{2} \ .
\end{equation}
Now, since $e^{\phi} \underset{\tau \sim 0}{\to} - \frac{\pi}{2}\frac{1}{\tau_0}$, using \eqref{eq: N_eff} and \eqref{eq: M_eff}, we get that
\begin{equation}
    M_{eff} (\tau) \underset{\tau \sim 0}{\to} M_0, \qquad N_{eff} (\tau) \underset{\tau \sim 0}{\to} N_0  \ .
\end{equation}
This implies that in the deep IR, where the gauge theory confines, the effective number of regular D3-branes equals $N_0$, while $M_0$ has to be interpreted as the number of fractional D3-branes at the same scale. In section \ref{sec: field theory IR} a vacuum was singled out which sits at the origin of the mesonic branch and whose low energy effective theory is that of a single-node quiver, describing pure $USp$ SYM. This is the only vacuum whose dual is expected to be pure geometry, very much like the baryonic branch of the KS theory for $N= k M$. Setting $N_0=0$ implies that there are no regular D3-branes left in the IR but only fractional ones, 
matching the dynamics of the isolated vacuum. Therefore, from now on, we will set $N_0=0$.   
 
Similarly to \cite{Benini:2007gx}, there is in fact a curvature singularity at $\tau=0$, even though the metric looks smooth. This can be traced to the fact that the $g_{\tau\tau}$ metric component in \eqref{eq: metric deformed conifold}, among others, approaches the constant value displayed in \eqref{metricIR} with a subleading term which is linear in $\tau$. This has the effect of producing Riemann tensor (orthonormal) components that behave like ${\cal R} \sim 1/\tau$. In contrast, in the KS solution \cite{Klebanov:2000hb} the subleading terms are quadratic in $\tau$, yielding a completely regular Riemann tensor. Note that since the dilaton is finite at $\tau=0$, the singularity is present both in the Einstein and string frame. In \cite{Benini:2007gx,Bigazzi:2008qq} such singularity has been interpreted as an artifact of the smearing approximation. In our case, it is then natural to attribute the singularity to the presence of the smeared O7-plane source. As already observed, the smeared solution can be understood as a truncation to the zero modes of the would-be localized solution. This suggests that the singularity arises because of such truncation and, consequently, is not expected to have a dual gauge theory interpretation, consistently with the analysis presented in section \ref{sec:field theory analysis}.

One may ask how Wilson loops behave in our background.\footnote{See, e.g.~\cite{Sonnenschein:1999if, Sonnenschein:1999re} for a discussion of Wilson loops in confining holographic set-ups.} The expectation is to find an area law only for those charged under the (center) 1-form symmetry, which is $\mathbb{Z}_2$, and a perimeter law for the others. This suggests the existence of a string worldsheet which, when considered individually, exhibits an area law, but when taken with multiplicity two, instead follows a perimeter law. While we postpone finding such a string to a future work, we expect it to be related to the presence of a torsion cycle in the orientifolded geometry \cite{Witten:1998xy}.

\subsubsection{UV limit}
\label{sec: UV of the SUGRA solution}

At large $\tau$, we expect our (covering) geometry to approach the undeformed conifold, since the dual gauge theory hits a fixed point in the extreme UV. 
To see that the expectation from gauge theory is met by our supergravity solution, let us start by studying the asymptotic behavior of the RR fluxes for large $\tau$. Expanding \eqref{eq: solution of fluxes equation} one gets 
\begin{equation}
\label{UV funcitons}
    e^{\phi} \sim  \frac{\pi}{2\tau},\qquad e^{-\phi} f \sim e^{-\phi} k \sim \frac{\tau}{2}, \qquad F \sim \frac{1}{2} \ .
\end{equation}
This implies that $f \sim k \sim  \frac{\pi}{4}$, which in turn gives 
\begin{equation} \label{eq: Meff and Neff in the UV}
\begin{split}
    M_{eff} (\tau) &= M_0 \biggl(1 -\frac{2}{ \pi} (f + k) \biggr) \xrightarrow[\tau \rightarrow \infty]{} M_0 \biggl(1 -\frac{2}{\pi} (\frac{\pi}{2}) \biggr) = 0, \\
    N_{eff} (\tau) &= \frac{M_0^{\, 2}}{\pi} \biggl( f - (f - k) F - \frac{2}{ \pi} f k \biggr) \xrightarrow[\tau \rightarrow \infty]{} \frac{M_0^{\, 2}}{8} \ .
\end{split} \end{equation}
This shows that the effective D5-charge vanishes in the UV and that there is a quadratic relation between the number of fractional D3-branes in the IR and the number of regular D3-branes in the UV. This relation resembles the one between IR and UV gauge theory ranks, \eqref{eq: N0M0_1} and \eqref{eq: N0M0_2}: this connection will be sharpened in section \ref{sec: checks}. 

Let us now look at the asymptotic behavior of the metric. We are going to expand the different functions entering the metric to next to leading order around $\tau \sim \infty$. The expansion of $\tilde{\Lambda}$, eq.~\eqref{eq: Lambda 1}, gives
\begin{equation}
    \tilde{\Lambda}(\tau) \underset{\tau \to \infty}{\sim} 2 \tau^{\frac{1}{3}}e^{-\frac{1}{3}\tau}\left[1-\frac{2\tau_0+1}{6\tau}\right] \ ,
\end{equation}
and in turn, for the $G_i$ functions entering the ansatz \eqref{eq: metric Ansatz} and whose expressions can be found in eqs.~\eqref{eq: solution deformed conifold}
\begin{equation}
    e^{2G_1}\sim e^{2G_2}\sim \frac{1}{4} \mu^\frac{4}{3} \tau^\frac{1}{3}e^{\frac{2}{3}\tau}\left[1-\frac{1+2\tau_0}{6\tau}\right]\quad , \quad e^{2G_3} \sim \frac{3}{2} \mu^\frac{4}{3}  \tau^\frac{1}{3}e^{\frac{2}{3}\tau}\left[1+\frac{1-\tau_0}{3\tau}\right] \ .
\end{equation}
The UV limit is easier to analyze going back to the $r$ coordinate. To do so, we need to integrate equation \eqref{eq: definition of tau}, which gives 
\begin{equation}
\label{eq: expansion of the radius for large tau}
    r \sim \frac{\sqrt{3}}{\sqrt{2}} \mu^\frac{2}{3} \tau^\frac{1}{6}e^\frac{\tau}{3}\left[1-\frac{2+\tau_0}{6\tau}\right] \ .
\end{equation}
In terms of $r$, the $G_i$ functions can be expanded as 
\begin{equation}
 e^{2G_1}\sim e^{2G_2}\underset{r \to \infty}{\sim} \frac{r^2}{6}\left(1+\frac{1}{6\log{r}}\right) \quad , \quad
     e^{2G_3} \underset{r \to \infty}{\sim} r^2\left(1+\frac{1}{3\log{r}}\right) \ ,
\end{equation}
Notice that $g=1/\cosh \tau$ goes to 0 as $\tau\to\infty$. This implies that at leading order the metric is given by  
\begin{equation}
\label{UVmetric}
    ds^2_6\underset{r\to\infty}{\sim} [dr^2+\frac{r^2}{6}(\sigma_1^2+\sigma_2^2+\omega_1^2+\omega_2^2)+\frac{r^2}{9}(\sigma_3+\omega_3)^2] \ ,
\end{equation}
which is the metric of the undeformed conifold and hence the whole metric asymptotes to $AdS_5 \times T^{1,1}$. Higher order terms provide (inverse) logarithmic suppressed corrections. 

Let us look at the behavior of the warp factor in this regime. If we expand $H(x)$ defined in \eqref{def H(x)} at large $x$, we get 
 \begin{equation}
     H(x) \underset{x\to\infty}{\sim} -2x^{-\frac{2}{3}} e^{-\frac{4}{3}x}\left[1+\frac{2\tau_0+1}{3x}\right]
 \end{equation}
 so we find
 \begin{equation}\label{eq: UV expansion of h}
     h(\tau)\underset{\tau\to\infty}{\sim}\frac{\pi M_0^{\, 2}}{2 \mu^\frac{8}{3}}
    \int\limits_{\tau}^{\infty} dx \, x^{-\frac{2}{3}} e^{-\frac{4}{3}x}\left[1+\frac{2\tau_0+1}{3x}\right]\underset{\tau\to\infty}{\sim} \frac{3\pi M_0^{\, 2}}{8} \mu^{-\frac{8}{3}}\tau^{-\frac{2}{3}}e^{-\frac{4}{3}\tau}\left[1+\frac{4\tau_0-1}{6\tau}\right] \ . 
 \end{equation}
Finally, using \eqref{eq: expansion of the radius for large tau}, we find 
\begin{equation}
\label{eq: UV asymp warp factor}
     h(r)\underset{r\to\infty}{\sim} \frac{27\pi}{4r^4} \frac{M_0^{\, 2}}{8} \left[1-\frac{1}{2\log{r}}\right] \ ,
\end{equation}
which, up to  logarithmic-suppressed  corrections, is the warp factor of $AdS_5 \times T^{1,1}$. This shows that the supergravity solution asymptotes to pure $AdS_5 \times T^{1,1}$ geometry  with $\frac{M_0^2}{8}$ units of $F_5$ flux, in agreement with the existence of a UV fixed point in the dual field theory. As $r$ goes to $\infty$, however, $e^{\phi} N_{eff}$ goes to $0$, which means that the curvature in the string frame diverges (see, e.g.~\cite{Itzhaki:1998dd, Aharony:1999ti} for a discussion on this regime). Therefore, if we were to interpret the UV limit of our solution as some holographic superconformal field theory, it would be outside the regime in which classical supergravity is a good approximation. Still, in the following section we will show that for every choice of $M_0$ we can choose the integration constant $\tau_0$ such that the supergravity approximation remains valid up to parametrically large $r$. In particular, it can be a good description of the dual field theory at energies much larger than the scale at which the duality cascade stops.

\section{Gauge/gravity duality checks}
\label{sec: checks}

In this section we provide further checks for the duality between the orientifold quiver gauge theory presented in section \ref{sec:field theory analysis} and the supergravity solution of section \ref{sec: SUGRA solution}. In particular, we will show how the duality cascade, the running of the gauge couplings, the R-symmetry breaking pattern and the $a$ and $c$ anomalies are correctly captured by the supergravity background. Key to these matchings are the gauge/gravity relations\footnote{Note that, as compared to the flavored conifold \cite{Benini:2007gx}, there is an extra factor of 1/2 on the right hand side of both equations, due to the orientifold involution, which is a $\mathbb{Z}_2$ action on the covering space.}
\begin{equation}
\begin{split} 
\label{eq:gauge_couplings}
&\frac{8 \pi^{2}}{g_{1}^{2}}  + \frac{8 \pi^{2}}{g_{2}^{2}} = \pi e^{- \phi}  \quad , \quad 
 \frac{8 \pi^{2}}{g_{1}^{2}} - \frac{8 \pi^{2}}{g_{2}^{2}}  = 2 \pi e^{- \phi} \biggl[ \left( \frac{1}{4 \pi^2} \int_{S^{2}} B_{2} \right) \bmod 1 - \frac{1}{2} \biggr]  \ , 
\end{split}
\end{equation}
which can be easily obtained by expanding the D3-brane action in powers of the world-volume field strength ${\cal F}_{\mu\nu}$ on the supergravity background. It is worth stressing that, strictly speaking, eqs.~\eqref{eq:gauge_couplings} are only valid for the $\mathcal{N} = 2$ orbifold theory \cite{Grana:2001xn,Bertolini:2001qa}. However, as discussed e.g. in \cite{Strassler:2005qs}, they also hold for the conifold theory as long as $g_1 \simeq g_2$ and $e^{\phi} \gg 0$. In the UV, where $N$ is large, our $USp(N+M-2) \times USp(N)$ gauge theory differs from the Klebanov-Witten model only by $\mathcal{O} (M/N)$ corrections, which are suppressed. This means that we can trust \eqref{eq:gauge_couplings} for large $\tau$ (as long as we are not interested in the UV conformal manifold, where $1/N$ corrections are important, as argued in section \ref{sec: conformal manifold}). 
With that said, we now turn to a detailed discussion of the various gauge/gravity checks. 

\subsection{Seiberg duality cascade}
\label{sec: Siebergduality}

The effective number of regular and fractional D3-branes in our background is given by the formul\ae \eqref{eq: N_eff} and \eqref{eq: M_eff}, respectively.  Using the explicit form of the three functions $f$, $k$ and $F$ given in \eqref{eq: solution of fluxes equation}, one can easily see that $N_{eff}$ is a monotonically increasing function of the radial coordinate. Following \cite{Klebanov:2000nc,Klebanov:2000hb}, this means that the appropriate number of colors needed to describe our field theory is becoming smaller and smaller as we decrease the energy. $M_{eff}$ behaves in the opposite way, growing larger as $\tau$ decreases. This implies that the effective number of fractional D3-branes increases toward the IR. As already observed in section \ref{sec: UV of the SUGRA solution}, this is roughly the behavior of the RG flow of the  orientifold gauge theory, that was interpreted in terms of repeated Seiberg dualities. 

In what follows, we would like to make this agreement precise. To achieve this, we first need to make sense of how one can match the discrete shifts of the gauge group ranks under Seiberg duality with the continuous running of the supergravity fluxes, the Maxwell charges $N_{eff}$ and $M_{eff}$. 

The starting point is the observation that the $B_2$ flux is a periodic quantity in string theory. More specifically, $B_2$ is a gauge potential and its flux is not a gauge invariant quantity. This is made explicit in the relations \eqref{eq:gauge_couplings} by the $\bmod \, 1$ identification. Every time the $B_2$ flux is $0 \bmod 1$ one of the two couplings diverges. This is naturally interpreted as a breakdown of the effective description of the theory in terms of the original $USp(N^{(1)}) \times USp(N^{(2)})$ gauge group and the necessity of performing a Seiberg duality. Different values of $\tau$ for which that happens correspond to values of the $B_2$ flux related by a large gauge transformation. This suggests matching the ranks of the gauge groups with supergravity quantities at values of the holographic coordinate where the $B_2$ flux is an integer. 

As observed in \cite{Benini:2007gx}, when the flux of $B_2$ is an integer, the Maxwell charges $M_{eff}$ and $N_{eff}$ coincide with one of the gauge equivalent values of the corresponding Page charges \cite{Marolf:2000cb}, defined as
\begin{equation}
\label{Page chs}
    Q_{D5} = \frac{1}{4 \pi^{2}} \int_{S^{3}} dC_{2} \quad , \quad Q_{D3} =\frac{1}{(4\pi^2)^2}\int_{T^{1,1}} dC_{4} \ .
\end{equation}
These quantities are not gauge invariant but are topological and quantized. They can be easily computed recalling that
\begin{equation}\label{eq:potentials}
    dC_{2} = F_{3} - B_2 \wedge F_1, \quad dC_{4} = F_{5} - B_2 \wedge F_3 + \frac{1}{2} B_2 \wedge B_2 \wedge F_1 \ .
\end{equation}
In order to compare with the field theory analysis of section \ref{sec: full RG}, it is useful to start computing the charges in the deep IR, which corresponds to $\tau=0$, and run the cascade up toward the UV. From the ansatz \eqref{eq: ansatz three forms} and the explicit solution for the functions $f$ and $k$, eq.~\eqref{eq: solution of fluxes equation}, we see that $B_2 \rightarrow 0$ in this limit (in which a gauge choice for $B_{2}$ is assumed), which implies that the Page charges in the choice of gauge of the ansatz are 
\begin{equation}
    Q_{D5}^0 = M_{eff} (\tau =0) = M_0 \quad , \quad Q_{D3}^0 = N_{eff} (\tau =0) = 0 \ .
\end{equation}
Consider a large gauge transformation that sets $B_{2} = - \pi n \, \Omega_{2}$ (at $\tau=0$), with $n \in \mathbb{Z}$ and where
\begin{equation}
    \Omega_{2} = \frac{1}{2} (g^1 \wedge g^2 + g^3 \wedge g^4) \ = \frac{1}{2}(\sin{\theta_1}d\theta_1 d\varphi_1-\sin{\theta_2}d\theta_2 d\varphi_2)
\end{equation}
is the volume form of the non-trivial $S^2$ defined by $\theta_1=\theta_2$, $\varphi_1 = 2 \pi - \varphi_2$, with  $\tau$ and $\psi$ fixed \cite{Bertolini:2002yr}. Under this shift the Page charges \eqref{Page chs} 
become 
\begin{equation} \begin{split}
\label{eq: Page charges at step k}
    &Q_{D5}^n = Q_{D5}^0 + \frac{1}{4 \pi^2}\int_{S^3} n \pi \Omega_2 \wedge F_1 = M_0 - \frac{2 n}{4 \pi^2} \int_{S^3} \Omega_2 \wedge g^5 = M_0 - 4n , \\ 
    &Q_{D3}^n = Q_{D3}^0 + \frac{1}{(4 \pi^2)^2} \int_{T^{1,1}} n \pi \Omega_2 \wedge \biggl( dC_2 +\frac12 n \pi \Omega_2 \wedge F_1 \biggr) = n M_0 - 2n^2\; ,
\end{split}\end{equation}
where we have used the explicit form for $F_1$ given in \eqref{eq: F_1} and equations \eqref{eq:volume_5}, \eqref{eq:volume_3}. Going from $\tau=0$ to $\tau=\infty$, $M_{eff}$ and $N_{eff}$ change continuously from $M_0$ to $0$ and from $0$ to $M_0^2/8$, respectively, and hit many different gauge equivalent values of $D3$ and $D5$ Page charges. The $\tau_{n}$'s for which this happens is where we are supposed to match supergravity charges and field theory ranks. What supergravity is telling us is that the field theory description in terms of $N_n$ and $M_n$ is valid from the energy scale corresponding to $\tau_{n}$ to that corresponding to $\tau_{n+1}$, and that at those two points we need to go to a Seiberg dual description, {\it i.e.} perform a large gauge transformation on $B_2$. A comparison with the field theory analysis of section \ref{sec:field theory analysis} suggests the following identification between supergravity and gauge theory parameters
\begin{equation}\begin{split}\label{eq:ranks}
    &M_{eff} (\tau_{n}) = Q_{D5}^n = M_0 - 4n = N^{(1)}_n-N^{(2)}_n+2 = M_n \\
    &N_{eff} (\tau_{n}) = Q_{D3}^n = nM_0 - 2n^2 = N^{(2)}_n = N_n \ ,
\end{split}\end{equation}
where $N^{(1)}_n$ and $N^{(2)}_n$ are the number of colors for the two gauge groups at the $n^{th}$ step of the cascade going toward the UV. Eq.~\eqref{eq:ranks} precisely matches the ranks at each step and their shifts, given in \eqref{eq: shift at step k}.\footnote{Note that differently from the KS theory, the D5 Page charge does not simply correspond to the difference of the gauge group ranks. This is a generic feature, whenever the dilaton has a non-trivial profile, see e.g. \cite{Benini:2007gx,Benini:2007kg}.} 

Having just analyzed the running of the RR fluxes, we now want to discuss the range of validity of the supergravity description, connecting with the discussion at the end of section \ref{sec: UV of the SUGRA solution}. The supergravity approximation is accurate as long as $e^{\phi} N_{eff}\gg1$. In the deep IR, where confinement occurs, 
\begin{equation}
    e^{\phi} N_{eff} \sim  -\frac{\pi}{2} \frac{M_0}{\tau_0}\sim \frac{M_0}{|\tau_0|}\ ,
\end{equation}
so we need to ask that $\frac{M_0}{\abs{\tau_0}}\gg1$ for the supergravity approximation to be valid at the confinement scale. Note that if we want the string coupling to be small enough at the tip of the geometry, we also need to ensure that we have $|\tau_0|\gg1$, so that eventually $M_0\gg |\tau_0|\gg1$. In our model the dilaton runs and, as already noticed, at large $\tau$ the string frame curvature becomes large and the supergravity approximation must break down. We will now show that this breakdown occurs at a scale parametrically larger than the length of the cascade. To do this, we will compute the value of $\tau$ corresponding to the last step of the cascade and compare it to that for which $e^{\phi} N_{eff}$ becomes of order $1$.

First, expanding \eqref{eq: solution of fluxes equation} for large $\tau$ and neglecting exponentially suppressed terms, we get $f \sim \frac{\pi}{4} \bigl (1 - \frac{1-\tau_0}{\tau} \bigr)$. We then use this to write
\begin{equation}
    M_{eff} (\tau_\mathrm{last}) \sim M_0 \frac{1-\tau_0}{\tau_\mathrm{last}} = m \ ,
\end{equation}
where $m$ is the value of $M$ at the last step of the cascade (as explained before, it can be either $2$ or $4$). All we need is that $m\sim {\cal O}(1)$, so that we infer
\begin{equation}
    \tau_\mathrm{last}\sim M_0 |\tau_0|\ .
\end{equation}
Note that $\tau_\mathrm{last}$ is also a measure of the length of the cascade, which is therefore {\em long}, since $\tau_\mathrm{last}\gg1$. First of all, we can see if the last step of the cascade is still in the regime of validity of the supergravity approximation. We can check that at $\tau \sim \tau_\mathrm{last}$, we have $e^\phi N_{eff}\sim M_0^2/\tau_\mathrm{last} \sim M_0/|\tau_0|\gg1$, so that we are still very safe. In fact, one can easily check that $e^\phi N_{eff}$ stays the same order for $0<\tau<\tau_\mathrm{last}$.

We can then find the value $\tau_\mathrm{max}$ at which $e^\phi N_{eff}\sim 1$ and the supergravity approximation ceases to be valid. This is when
\begin{equation}
 e^\phi N_{eff} \sim \frac{M_0^2}{\tau_\mathrm{max}} \sim 1\ ,    
\end{equation}
which means 
\begin{equation}
    \tau_\mathrm{max} \sim M_0^2\ .
\end{equation}
It is now very easy to show that 
\begin{equation}
    \frac{\tau_\mathrm{max}}{\tau_\mathrm{last}} \sim \frac{M_0}{|\tau_0|}\gg 1\ ,
\end{equation}
which means that we can trust the solution to describe the entire cascade. In fact, the supergravity solution will remain reliable also in the range  $\tau_\mathrm{last} < \tau < \tau_\mathrm{max}$, which is a much wider energy range than that in which the cascade takes place. We interpret the behavior of the solution in this region as describing the very slow flow toward the field theory conformal manifold. Finally, at $\tau \sim \tau_\mathrm{max}$ the supergravity approximation breaks down, meaning that the solution cannot properly describe the field theory UV fixed point, as anticipated at the end of section \ref{sec: UV of the SUGRA solution}. 

\subsection{Couplings and \texorpdfstring{$\beta$}{b} functions} 
\label{sec: beta functions}

\begin{figure}[h!]
    \centering
    \begin{tikzpicture}
        \begin{scope}
            \clip (5,0) rectangle (13,6);
            \node[anchor=south west, inner sep=0] (full) at (5,0)
          {\includegraphics[width=0.6\textwidth]{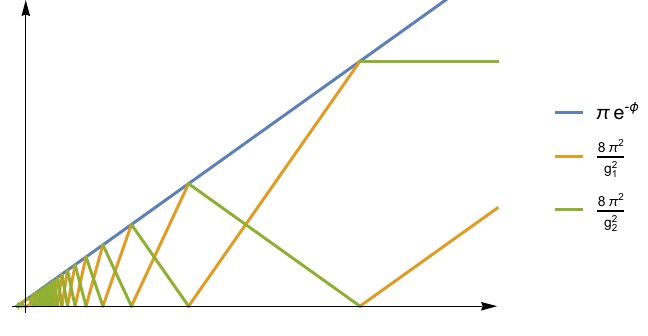}};
        \end{scope}
        \draw [line width = 1.2pt, bluewolfram] (14,3.5) -- (14.5,3.5);
        \node at (15.1,3.58) {$\pi e^{-\phi}$};
        \draw [line width = 1.2pt, orangewolfram] (14,2.7) -- (14.5,2.7);
        \node at (14.95,2.66) {$\frac{8\pi^2}{g_1^2}$};
        \draw [line width = 1.2pt, greenwolfram] (14,1.9) -- (14.5,1.9);
        \node at (14.95,1.86) {$\frac{8\pi^2}{g_2^2}$};
    
        \node[anchor=south west, inner sep=0] (zoom) at (-1,2.0)
          {\includegraphics[width=0.3\textwidth]{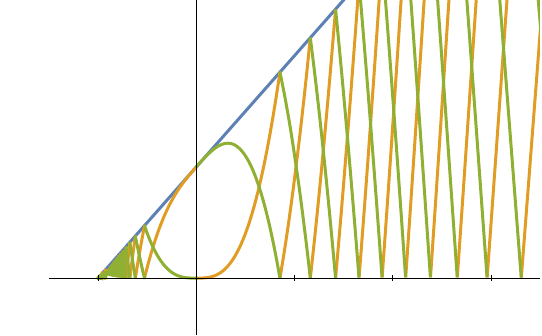}};
    
        \draw[line width=1 pt] (4.8,0.7) .. controls (3.6,1.0) and (3,1.5) .. (2.8,2.0); 
        \draw[-{Latex[length=2.5mm]}, thick] (2.81,1.95) -- +(-0.1,0.28);
        \draw[red, thick, dashed] (5,0.1) rectangle (6,0.8);
    
        
        \node at (10.75,0.2) {$\tau_{\text{last}}$};
        \node at (12.75,0.2) {$\tau$};
        \node at (-0.170, 2.3) {$\tau_0$};

    \end{tikzpicture}
    \caption{The running of gauge couplings, $8 \pi^{2}/g_{1}^{2}$ and $8 \pi^2/g_{2}^{2}$ for  $M_0=80$. A similar plot would hold choosing $M_0 = 2 \,\text{mod}\, 4$, the only qualitative difference being that at scales larger than $\tau_{\text{last}}$ both $\beta$ functions would be positive, see eq.~\eqref{eq: beta functions M=0}. The blue line corresponds to their sum, which diminishes toward the IR, exactly matching the field theory analysis of section \ref{sec:field theory analysis}. The supergravity approximation remains valid up to $\tau \sim\tau_{\text{max}} \gg \tau_{\text{last}}$, which lies far to the right, well beyond the range of $\tau$ shown in the figure. While in the UV the cascade stops, in the IR a duality wall shows up. However, as discussed in sections \ref{sec: field theory IR} as far as field theory analysis and \ref{sec: IR limit of SUGRA} for its supergravity dual, the duality wall is screened by confining dynamics, that the perturbative running of the gauge coupling displayed here, however, cannot grasp. The zoomed-in part shows how the cascade looks distorted in the deep IR. Yet, confinement occurs at a larger scale and excises that region.}
    \label{fig:cascade_plot}
\end{figure}
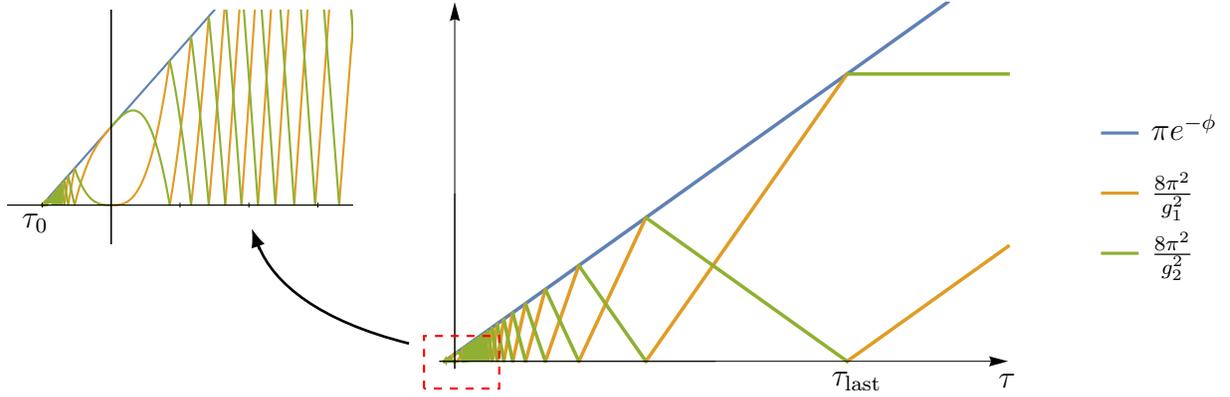

Elaborating on what we discussed above, we now want to match the gauge theory $\beta$ functions with those that one can compute using the holographic relations \eqref{eq:gauge_couplings}. The couplings $g_1$ and $g_2$ in those equations are not to be thought of as the same couplings all along the flow. In a given energy range $(\tau_n, \tau_{n+1})$, the couplings are those of the effective description $USp(N_n + M_n-2) \times USp(N_n)$ discussed in the previous section.
For large $\tau$, using \eqref{eq: ansatz three forms}, we get
\begin{equation}\label{eq:asymptotic_B}
    \frac{1}{4 \pi^2} \int_{S^{2}} B_{2} \underset{\tau \to \infty}{\sim}  \frac{M_0}{\pi} f(\tau) \ .
\end{equation}
From \eqref{eq:gauge_couplings} and \eqref{eq:asymptotic_B} we can then determine the gauge couplings as a function of the radial coordinate in the UV as
\begin{equation}\label{eq:running}
    \frac{8 \pi^{2}}{g_{1}^{2}} \underset{\tau \to \infty}{\sim} \pi e^{-\phi} \biggl[ \biggl( \frac{M_0}{\pi} f(\tau) \biggr)\bmod 1 \biggr] \quad , \quad  \frac{8 \pi^{2}}{g_{2}^{2}} \underset{\tau \to \infty}{\sim} \pi e^{-\phi} \biggl[ \biggl(- \frac{M_0}{\pi} f(\tau) \biggr)\bmod 1 \biggr] .
\end{equation}
where the asymptotic expression for $e^{-\phi} f$ is given in eq.~\eqref{UV funcitons} and goes as $\tau/2$ to leading order. The resulting plot, which shows the gauge couplings as a function of $\tau$, is depicted in figure  \ref{fig:cascade_plot} and displays the characteristic cascading behavior.

In order to match the field theory and supergravity expressions for the $\beta$ functions, we need an energy/radius relation between the field theory scale $\mu$ and the holographic coordinate, which we can state as 
\begin{equation}\label{eq: energy-radius relation}
    \log \mu = l(\tau), 
\end{equation}
where $\mu$ is the RG scale. A way to find $l(\tau)$ is to impose, e.g., the matching of the $\beta$ function for the sum of the gauge couplings, and then check that $\beta_1$ and $\beta_2$ are also matched. This is particularly convenient, since the $\beta$ function for $8 \pi^{2}/g_{1}^{2} + 8 \pi^{2}/g_{2}^{2}$ is constant in the UV up to $\mathcal{O}(1/N)$ corrections. More precisely, all effective descriptions in terms of $USp(N_n + M_n - 2) \times USp(N_n)$ gauge groups have $\beta_+ \simeq 6$, as long as the anomalous dimension of the chiral superfields is close to $\gamma = - \frac{1}{2}$ (which is true at high enough energies). 

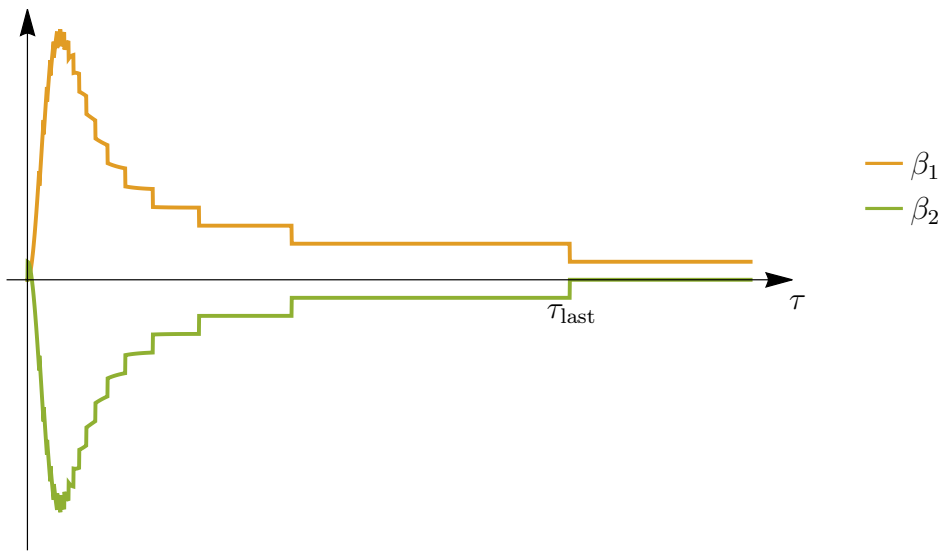
\begin{figure}[h!]
    \centering
    \begin{tikzpicture}
        \begin{scope}
            \clip (0,0) rectangle (11,8);
            \node[anchor=south west, inner sep=0] (full) at (0,0)
          {\includegraphics[width=0.8\textwidth]{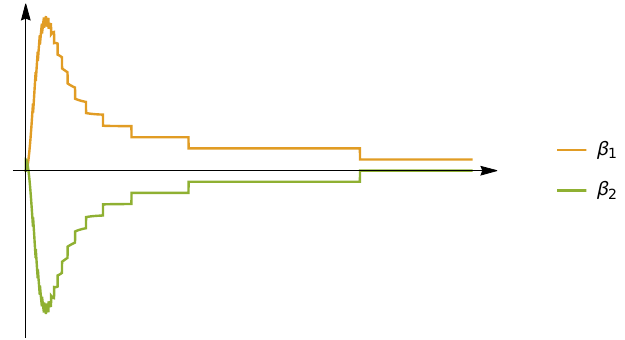}};
        \end{scope}
        \draw [line width = 1.2pt, orangewolfram] (11.7,5.2) -- (12.2,5.2);
        \node at (12.5,5.16) {$\beta_1$};
        \draw [line width = 1.2pt, greenwolfram] (11.7,4.6) -- (12.2,4.6);
        \node at (12.5,4.56) {$\beta_2$};
        \node at (7.8,3.2) {$\tau_{\text{last}}$};
        \node at (10.8,3.35) {$\tau$};
    \end{tikzpicture}
    \caption{ Supergravity prediction for the gauge theory $\beta$-functions. At large $\tau$, they exhibit the discontinuous logarithmic running characteristic of cascading theories. Also in this case, $\tau_{\text{max}}$ lies on the far right of the graph, outside of the depicted region.}
    \label{fig:beta_plot}
\end{figure}

From eqs.~\eqref{eq:gauge_couplings} and \eqref{eq: diffeq phi} we have 
\begin{equation}
    \beta_{+} \equiv \beta_{1} + \beta_{2} = \frac{\pi}{l'(\tau)} \frac{d}{d \tau} e^{- \phi} = \frac{2}{l'(\tau)} \ ,
\end{equation}
which, imposing that $\beta_+=6$, fixes $l'(\tau) = \frac{1}{3}$ and in turn 
\begin{equation}\label{eq:beta_derivative}
    \frac{d}{d \log \mu} = 3 \frac{d}{d \tau} \ .
\end{equation}
For large $\tau$, from \eqref{eq: expansion of the radius for large tau} we find that $ \tau \sim 3 \log r$, meaning that $r$ is proportional to the RG scale $\mu$. Eqs.~\eqref{eq:running} can be written as
\begin{equation}
\begin{split}
\label{eq:couplings_mod}
\frac{8 \pi^{2}}{g_{1}^{2}} \underset{\tau \to \infty}{\sim} M_0 e^{-\phi} f - n \pi e^{-\phi} \quad , \quad
\frac{8 \pi^{2}}{g_{2}^{2}} \underset{\tau \to \infty}{\sim}  - M_0 e^{-\phi} f + (n+1) \pi e^{-\phi} \ ,
\end{split}
\end{equation}
where $n =  \left \lfloor {\frac{M_0}{\pi} f(\tau)} \right \rfloor$ is the step of the cascade at which $g_1$ and $g_2$ have to be interpreted as the couplings of the effective description valid for $\tau \in (\tau_n, \tau_{n+1})$.
Using \eqref{eq:couplings_mod} and \eqref{eq:beta_derivative} we can finally compute the $\beta$ functions and get 
\begin{equation}\begin{split}\label{eq:cascade_plot}
    &\beta_1 \sim 3 \frac{d}{d \tau} \frac{8 \pi^2}{g_1^2} \sim  3 \frac{d}{d \tau} \biggl( M_0 e^{-\phi} f - n \pi e^{-\phi} \biggr)  \\ &\beta_2 \sim 3 \frac{d}{d \tau} \frac{8 \pi^2}{g_1^2} \sim  3 \frac{d}{d \tau} \biggl( - M_0 e^{-\phi} f + (n+1) \pi e^{-\phi} \biggr)  \ .
\end{split}\end{equation}
In figure \ref{fig:beta_plot} we plot $\beta_1$ and $\beta_2$ as functions of $\tau$. Going toward the UV $\beta_1$ and $\beta_2$ become step functions, the points of discontinuity being the values of $\tau$ where Seiberg duality should be done and a different effective description takes over. This means that for large $\tau$ the RG flow will be logarithmic in each energy range. The result is in agreement with \eqref{eq: beta functions}, where if we neglect $\delta_0$ we get $\beta$ functions that are constants and depend on the ranks of the gauge groups. Performing a Seiberg duality means changing the gauge groups; therefore \eqref{eq: beta functions} predicts a step function behavior for $\beta_1$ and $\beta_2$, which is exactly what we find here.

In fact, recalling that the leading asymptotic at large $\tau$ are $e^{-\phi} \sim \frac{2}{\pi} \tau$ and $e^{-\phi} f \sim \frac{\tau}{2}$, we can precisely match the different heights of the UV step functions in the different energy ranges
\begin{equation}\begin{split}
\label{beta sugra}
    &\beta_1 \sim 3 \frac{d}{d \tau} \biggl( \frac{M_0 \tau}{2} - 2 n \tau  \biggr) \sim \frac{3}{2} (M_0-4n) = \frac{3}{2} M_n  \\ &\beta_2 \sim 3 \frac{d}{d \tau} \biggl(- \frac{M_0 \tau}{2} + 2 (n+1) \tau  \biggr) \sim -\frac{3}{2} (M_0-4n)+6 = - \frac{3}{2} (M_n - 4) \ ,
\end{split}\end{equation}
where we have used the relations \eqref{eq:ranks} to express the result in terms of the number of colors of the $USp(N_n + M_n-2) \times USp(N_n)$ effective description. Eqs.~\eqref{beta sugra} exactly reproduce the gauge theory computation \eqref{eq: beta functions}. 

Let us conclude by commenting on the relation between the field theory and its gravity dual in the extreme UV. We have argued in section \ref{sec:field theory analysis} that the gauge theory flows from a (one-dimensional) conformal manifold. This is specified by one coupling taken to zero, which is $\eta$ for $M_{UV}=2$ and $g_1$ for $M_{UV}=4$, see eqs.~\eqref{an1} and \eqref{an2}. We take this zero-coupling sector of the UV SCFT to be the reason why the supergravity dual eventually breaks down in this regime, in the sense that the curvature becomes large in string units. Indeed, if we take the example of fig.~\ref{fig:cascade_plot}, which refers to the situation $M_{UV}=4$, we see that the coupling $g_1$ is asymptotically free. The case of $M_{UV}=2$ is more subtle, since generically both gauge couplings are supposed to be eventually fixed to a finite value. This means that in this case the modification with respect to the gravity answer is sharper, since also the sum of their inverse squares (the blue envelope in fig.~\ref{fig:cascade_plot}) is expected to flatten to a constant. This is not captured by our solution, so that we expect this behavior to occur after the value $\tau_\mathrm{max}$ at which the curvature becomes stringy. Nonetheless, the fact that in both cases the geometry asymptotes to $AdS_5 \times T^{1,1}$ at large $r$ is in nice agreement with the expected large anomalous dimensions of the matter fields at the conformal fixed point.

\subsection{R-symmetry breaking pattern}
\label{sec: R-symmetry}

The undeformed conifold has $SU(2)\times SU(2) \times U(1)$ isometry group. The $U(1)$ factor, which corresponds to the classical R-symmetry in the dual gauge theory, is generated by shifts of the fiber coordinate $\psi$ \cite{Kehagias:1998gn}. In the cascading gauge theory only a discrete subgroup of $U(1)_R$ is a symmetry at the quantum level. In supergravity the quantum breaking of the classical symmetry is  due to the non invariance of the RR potentials $C_2$ and $C_0$, while the leftover discrete R-symmetry is encoded in their monodromies \cite{Bertolini:2001qa,Klebanov:2002gr,Bertolini:2002xu}, as we now explain. 

In the supergravity solution, all gauge invariant fields, including the metric, are invariant under shifts of $\psi$ (we focus on the UV limit \eqref{UVmetric}, which is the relevant one to capture the field theory anomaly). In general, gauge potentials defined locally on each patch do not possess such a symmetry, so the $U(1)$ symmetry associated to $\psi$ rotations may look fully broken by the solution. However, gauge potentials are not gauge invariant quantities, so any shift of $\psi$ that maps a gauge potential to a gauge equivalent configuration is still a symmetry of the background. In determining such redundant shifts, there are two subtleties we should take care of in our case. First, due to the smearing procedure that we needed to perform to get the solution, $F_1$ is not closed. In order to define a $C_0$, following \cite{Benini:2007gx}, we restrict to a submanifold $\mathcal{C}_4$ inside the deformed conifold, specified by $\theta_1=\theta_2=\theta$, $\varphi_1=2\pi-\varphi_2=\varphi$ while being along the radial and fiber coordinate $\psi$, where $dF_1=0$. We therefore define
\begin{equation}\label{eq: effective field strengths}
    F_1^{eff} \equiv F_1 \rvert_{\mathcal{C}_4}=-\frac{2}{\pi}d\psi \ .
\end{equation}
Conversely, $C_2$ can always be defined as $dC_2=F_3-B_2\wedge F_1$. So, on ${\cal C}_4$ we can define the RR potentials as
\begin{equation}\label{eq: effective RR potentials}
    C_0^{eff}=-\frac{2}{\pi}\psi \quad, \quad C_2^{eff}=\frac{M_0}{2} \psi\, \sin{\theta}\,d\theta\wedge d\varphi~.
\end{equation}
The second subtlety is related to the quantization of the large gauge transformations of these potentials: due to orientifold action, when we compute integrals on the covering space, both D5 and D7 charges get a factor of $2$, hence modifying the quantization condition of the relative gauge potentials. Explicitly, with a standard Dirac monopole argument, one gets  
\begin{align}
\label{eq: large gauge C0}
    C_0^{eff}&\to C_0^{eff}+\lambda_0 ,  &\lambda_0=2k_0, \quad k_0 \in \mathbb{Z} \ , \\
\label{eq: large gauge C2}
     C_2^{eff}&\to C_2^{eff}+\lambda_2^c,  &\lambda_2^c=2\pi\Omega_2 k_2^c, \quad k_2^c \in \mathbb{Z} \ .
\end{align}
where $\Omega_2$ is the volume form on the $S^2$. In \eqref{eq: large gauge C2} we have omitted the shift of $C_2^{eff}$ that follows from a large gauge transformation of $B_2$, as this is not relevant for the present discussion. 
From \eqref{eq: large gauge C0} and \eqref{eq: large gauge C2} we see that the shifts of $\psi$ which map $C_0^{eff}$ and $C_2^{eff}$, \eqref{eq: effective RR potentials}, to gauge equivalent configurations are
\begin{equation}
\begin{split}
    C_0^{eff}(\psi+\alpha)&=C_0^{eff}(\psi)+\lambda_0 \rightarrow \alpha= \pi k_0^c, \quad k_0^c=0,1,2,3 \\
    C_2^{eff}(\psi+\alpha)&=C_2^{eff}(\psi)+\lambda_2^c \rightarrow \alpha= \frac{4\pi}{M_0}k_2^c, \quad k_2^c=0,1,...,M_0-1
\end{split}
\end{equation}
This implies the following anomalous breaking of the R-symmetry predicted by the supergravity solution 
\begin{equation}
\label{R-sym breaking 1}
    U(1)\to \mathbb{Z}_{gcd(M_0,4)}~,
\end{equation}
in exact agreement with the field theory result \eqref{eq: ft nonanR}. 

One can obtain the same result thinking in terms of gauge invariant charged operators, by considering the Wess-Zumino couplings of the D$(-1)$ and D1-branes. The corresponding charged operators are
\begin{equation}
    W_{0,p}(x) = e^{2\pi i p \, C_0^{eff}} \quad , \quad W_{2,q} (S^2) = e^{ \frac{iq}{2\pi} \int_{S^2} C_2^{eff}} \ .
\end{equation}
Under a shift of $\psi \rightarrow \psi+\alpha$ they transform as 
\begin{align}
\label{Wop0}
    & W_{0,p}(x) \rightarrow W_{0,p}(x) e^{-4ip\alpha}\\
\label{Wop2}
    & W_{2,q} (S^2) \rightarrow W_{2,q} (S^2) e^{iq M_0 \alpha} \ .
\end{align}
Because of the orientifold projection, the minimal D$(-1)$ and D1 charges are half of those on the conifold (this is equivalent to the fact that the minimal magnetic charges, {\it i.e.} D5- and D7-brane charges,  equal $2$). This implies that $p,q \in \mathbb{Z}/2$ and their minimal value is $1/2$. Imposing that the $\psi$ shift leaves both operators invariant gives
\begin{equation}
    2\alpha = 0 \mod 2\pi \quad , \quad \frac{M_0\alpha}{2} = 0 \mod 2\pi \ .
\end{equation}
Since $\psi \in [0,4\pi)$ we conclude that the $U(1)_R$ symmetry is broken to 
\begin{equation}
    U(1)\to \mathbb{Z}_{gcd(M_0,4)} \ ,
\end{equation}
which is the same as \eqref{R-sym breaking 1}. 

As discussed in section \ref{sec: field theory IR}, the field theory vacuum we want to describe is a confining one. The left over pure $USp$ SYM at the bottom of the cascade confines and the gaugino bilinear condenses. This spontaneously breaks the R-symmetry to $\mathbb{Z}_2$, independently on whether $M_0 = 0 \,\text{mod} \,4$ or $M_0= 2 \,\text{mod} \,4$. Unlike the anomalous breaking, which relies on the large $\tau$ regime of the supergravity solution, this breaking is a IR effect and it is hence expected to be captured by the full background whose internal geometry, as discussed in section \ref{sec: IR limit of SUGRA}, is that of the deformed conifold, to leading order in $\tau$. Unlike the singular conifold, the deformed conifold metric is not invariant under arbitrary shifts in $\psi$, but actually preserves only a $\mathbb{Z}_2$ subgroup \cite{Klebanov:2000hb}. So the full solution presented in section \ref{sec: SUGRA solution} preserves only a $\mathbb{Z}_2$ R-symmetry, in agreement with field theory predictions.

\subsection{Holographic central charge}
\label{sec: a charge}

One further check we want to provide regards the holographic computation of the $a$ and $c$ charges of the superconformal field theory which UV-completes the orientifold gauge theory.  

In a ${\cal N}=1$ SCFT the $a$ and $c$ anomalies are related to the superconformal R-charge anomaly as
\begin{equation}
a = \frac{3}{32} \left( 3 \text{Tr} R^3 -  \text{Tr} R \right) \quad , \quad 
c = \frac{1}{32} \left( 9 \text{Tr} R^3 -  5 \text{Tr} R \right)\ .
\end{equation}
A simple but lengthy computation gives for the two (classes of) SCFTs which UV-complete the orientifold cascade, $USp(N_{\text{UV}})\times USp(N_{\text{UV}})$ for $M_{\text{UV}}= 2$ and $USp(N_{\text{UV}}+2)\times USp(N_{\text{UV}})$ for $M_{\text{UV}}= 4$, respectively
\begin{itemize}
    \item $M_{\text{UV}} = 2$: 
    \begin{equation}
    a = \frac{3}{32} \left( \frac{9}{4} N_{\text{UV}}^2 - \frac{1}{2} N_{\text{UV}} - 9 - \frac{6}{N_{\text{UV}}} \right)  \; , \quad 
c = \frac{1}{32} \left( \frac{27}{4} N_{\text{UV}}^2 + \frac{1}{2} N_{\text{UV}} - 27 - \frac{18}{N_{\text{UV}}} \right)\ .
    \end{equation}

    \item $M_{\text{UV}} = 4$: 
    \begin{equation}
    a = \frac{3}{32} \left( \frac{9}{4} N_{\text{UV}}^2 + \frac{13}{2} N_{\text{UV}} +6 \right)  \quad , \quad 
c = \frac{1}{32} \left( \frac{27}{4} N_{\text{UV}}^2 + \frac{35}{2} N_{\text{UV}} + 12 \right)\ .
    \end{equation}
    \end{itemize}
As for any SCFT that can admit a holographic dual, $a$ and $c$ consistently agree at leading $N_{\text{UV}}^2$ order. This is all supergravity predictions can be compared with. Subleading contributions to $a$ and $c$ are expected to come from string loop corrections, corresponding to higher derivative corrections in the supergravity effective action, proportional to ${\cal R}^2$ and ${\cal R}^4$ terms, and are typically $1/N_{\text UV}^2$ suppressed with respect to the leading contribution \cite{Henningson:1998gx,Anselmi:1998zb}. The expressions above also include contributions suppressed by $1/N_{\text UV}$, only. These are expected to originate from the orientifold, much like what occurs in, e.g., ${\cal N}=4$ SYM with symplectic or orthogonal gauge groups. Notice, further, that the leading-order expressions for $a$ and $c$ in the two cases, $M_{\text{UV}} = 2$ and $M_{\text{UV}} = 4$, are the same: for both SCFT at leading order 
\begin{equation}
\label{ac_qft}
    a=c=\frac{27}{128} N_{\text{UV}}^2 \ .
\end{equation}
The holographic dictionary relates the $a$ charge with the volume of the internal manifold $X_5$ via the relation \cite{Henningson:1998gx,Gubser:1998vd,Anselmi:1998zb}
\begin{equation}
\label{ac_hol}
    a = \frac{N^2}{4} \frac{\pi^3}{V(X_5)} \ ,
\end{equation}
where $N$ are the units of $F_5$ flux (equivalently, the number of D3-branes) and $\pi^3$ the volume of the unit 5-sphere. To compare \eqref{ac_hol} with the field theory answer, we recall that our solution,  which asymptotes to $AdS_5 \times T^{1,1}$ (see eq.~\eqref{eq: UV asymp warp factor} and the discussion therein), was derived in the covering space. In the physical space, the volume of the internal space $X_5$ is half that of $T^{1,1}$, that is $V(X_5)= \frac{8}{27} \pi^3$, and $N= \frac{1}{2} N_{eff}$  (see, e.g.~\cite{Antinucci:2020yki, Antinucci:2021edv}) with $N_{eff} = M_0^2/8$ given by eq.~\eqref{eq: Meff and Neff in the UV}. Plugging these numbers into eq.~\eqref{ac_hol} we get 
\begin{equation}
\label{ac_hol2}
    a = \frac{27}{128} \frac{M_0^4}{64} 
    \ ,
\end{equation}
which, upon using eqs.~\eqref{eq: N0M0_1} and \eqref{eq: N0M0_2}, exactly matches the field theory prediction \eqref{ac_qft}.

While central charges can be rigorously defined only at conformal fixed points, it is nevertheless insightful to consider their generalizations as functions—commonly referred to as central functions—which interpolate between the UV and IR regimes of the gauge theory \cite{Cardy:1988cwa,Komargodski:2011vj}. This idea was first proposed in the holographic context in \cite{Girardello:1998pd, Freedman:1999gp}. The goal is to define a quantity that matches the central charge of the UV CFT at high energies and decreases monotonically along the RG flow. In cases where the IR theory is gapped, as in our setup, we expect the central function to vanish in the deep IR, reflecting the absence of massless degrees of freedom in the confining vacuum. Such a function can be interpreted as a measure of the number of effective degrees of freedom along the flow, making its potential monotonicity particularly compelling. 

\begin{figure}[h!]
  \centering
  \begin{tikzpicture}[scale=0.9, transform shape]
    \node[anchor=south west, inner sep=0] (full) at (5,0)
      {\includegraphics[width=0.55\textwidth]{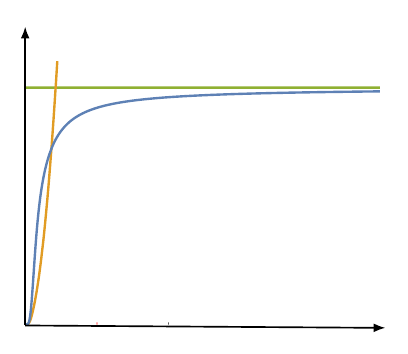}};

    \node[anchor=south west, inner sep=0] (zoom) at (-1.5,2.0)
      {\includegraphics[width=0.35\textwidth]{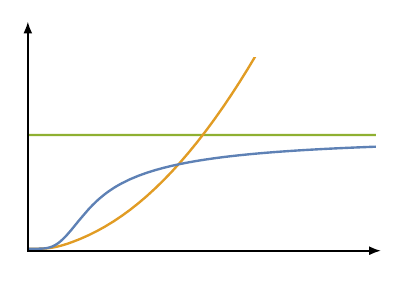}};

    \draw[line width=1 pt] (4.8,1) .. controls (3.6,0.8) and (3,1.3) .. (2.8,1.7); 
    \draw[-{Latex[length=2.5mm]}, thick] (2.8,1.7) -- +(-0.1,0.18);
    \draw[red, thick, dashed] (5.4,0.5) rectangle (7.206,6.4);

    

    \draw [line width = 1.2pt, bluewolfram] (14.7,5.2) -- (15.2,5.2);
    \node at (15.8,5.2) {$c(\tau)$};
    \draw [line width = 1.2pt, orangewolfram] (14.7,4.6) -- (15.2,4.6);
    \node at (16,4.65) {$c_{\text{KS}}(\tau)$};
    \draw [line width = 1.2pt, greenwolfram] (14.7,4) -- (15.2,4);
    \node at (15.75,4.0) {$c_{\text{UV}}$};
    \draw[thick] (8.84,0.53) -- (8.84,0.71);
    \node at (8.9, 0.37) {$\tau_{\text{last}}$};
    \node at (13.8,0.3) {$\tau$};
        
  \end{tikzpicture}
   \caption{The holographic $c(\tau)$ function \eqref{holc0}, blue line in the figure. It saturates to the central charge \eqref{ac_qft} as $\tau \to \infty$ and goes to zero in the IR, in agreement with a RG flow from a UV fixed point to a confining vacuum. As in previous figures, $\tau_{\text{max}}$ lies outside  the depicted region. The zoomed-in part shows the behavior at small values  of $\tau$. The yellow line describes the same quantity evaluated in the KS background, fixing the overall constant so that the small $\tau$ behavior coincides with that of our solution. For large $\tau$, $c_{\text{KS}}(\tau)$ diverges, consistently with the UV completion involving an infinite number of degrees of freedom. In the IR, however, it grows more slowly than ours as one climbs up the cascade. This is because the cascade steps of the orientifold gauge theory are more dense near $\tau =0$, see figure \ref{fig:cascade_plot}.}
   \label{fig:c(r)}
\end{figure}
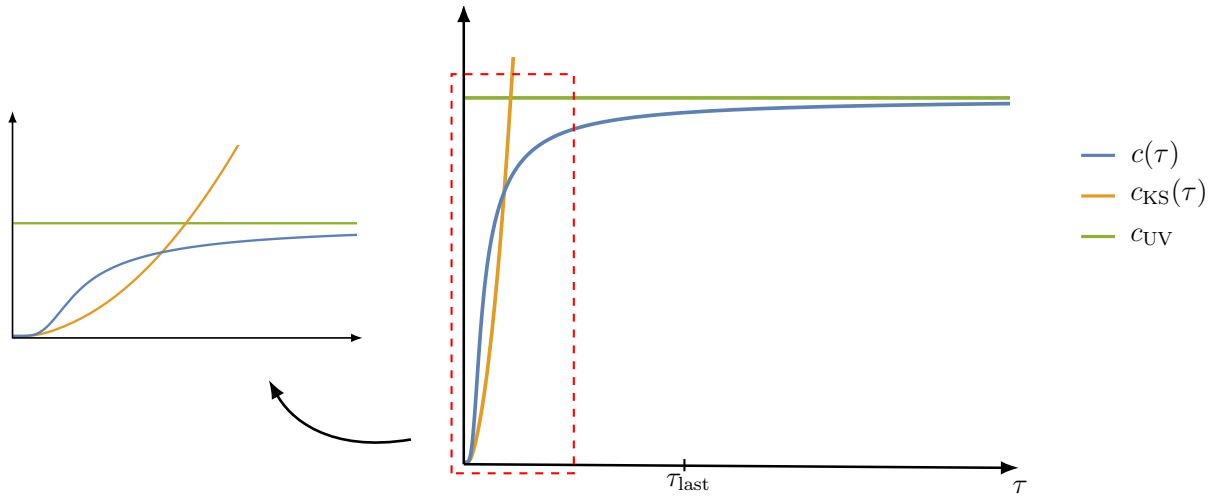

The holographic $c$-function is a proper extension of \eqref{ac_hol} which depends on the warp factor $h(\tau)$ and on the internal volume being now a function of the holographic coordinate, as dictated by the metric \eqref{eq: metric Ansatz}, that is $V_\tau(X_5) = \int \prod_{i=1}^5 d y^i \sqrt{{ \det g_{ij}}}$ with $y^i=(\theta_1,\theta_2,\varphi_1,\varphi_2,\psi)$. Following the conventions of \cite{Macpherson:2014eza}, to which we refer for details, fixing the overall normalization as to match the UV result  \eqref{ac_hol2}, the precise expression one gets for the holographic $c$-function is
\begin{equation}
\label{holc0}
    c(\tau) = \frac{4}{3 \pi^2} \frac{e^{2 G_1(\tau)+2 G_2(\tau)+ 4 G_3(\tau)} h^2(\tau)}{\left(4 G_1'(\tau)+4 G_2'(\tau)+2 G_3'(\tau) +h'(\tau)/h(\tau)\right)^3} \ .
\end{equation}
Expanding the above expression at large and small $\tau$, one finds that it correctly interpolates between a constant value in the UV, matching \eqref{ac_hol2}, and zero in the IR, as expected. This behavior is illustrated in figure \ref{fig:c(r)}, where the monotonicity of the $c$-function is also manifest.

It is instructive to compare our results with those   one gets for the KS solution \cite{Klebanov:2000hb}, see \cite{Elander:2011mh} for an explicit derivation. Also in that case, the $c$-function \eqref{holc0} vanishes in the IR, reflecting the fact that the vacuum on the baryonic branch is gapped (except for the Goldstone chiral superfield  of the spontaneously broken baryonic symmetry, whose contribution is not captured in the supergravity limit). However, in contrast to our setup, it  diverges quadratically in $\tau$ in the UV, consistently with the expected UV completion of the KS theory involving an infinite number of degrees of freedom. All these features are displayed in figure \ref{fig:c(r)}.

\section*{Acknowledgments}

We thank Francesco Benini for enlightening comments and exchange of ideas. We also thank Daniele Migliorati and  Valdo Tatitscheff for useful discussions. Finally, we are grateful to Igor Klebanov and Carlos Nunez for helpful comments on the draft. F.A., M.B., E.G.-V. and P.M. are supported by MIUR PRIN Grant 2020KR4KN2 “String Theory as a bridge between Gauge Theories and Quantum Gravity” and by INFN Iniziativa Specifica ST\&FI. R.A. is a Research Director of the F.R.S.-FNRS (Belgium). The research of R.A. is supported by IISN-Belgium (convention 4.4503.15) and through an ARC advanced project.

\appendix
\section{Solution of the equations of motion}
\label{app: SUGRA solution}

In this appendix, we solve the type IIB supergravity equations of motion, starting from the ansatz discussed in section \ref{sec: ansatz}. As usual for supersymmetric solutions as ours, it is enough to solve the first order BPS equations, since any such solution is also a solution of the equations of motion (this was explicitly checked for the case at hand in \cite{Benini:2006hh}). 

The BPS equations that the solution should satisfy imply that there exist two such classes, which differ by the way the following constraint between the functions appearing in the metric ansatz  \eqref{eq: metric Ansatz}, namely 
\begin{equation}
\label{eq: g constraint}
g \left[g^2 -1 + e^{2(G_1-G_2)}\right]= 0 \ , 
\end{equation}
is satisfied. As observed in \cite{Benini:2007gx}, solutions satisfying \eqref{eq: g constraint} by $g=0$ belong to Klebanov-Tseytlin-like backgrounds \cite{Klebanov:2000nc}, while those satisfying \eqref{eq: g constraint} by $g^2 = 1 - e^{2(G_1-G_2)}$ provide confining Klebanov-Strassler-like backgrounds \cite{Klebanov:2000hb}. Since our gauge theory confines in the IR, we impose the supergravity solution should satisfy the latter. 

With this choice, the differential equations that the functions $G_1,G_2,G_3$ appearing in the metric  \eqref{eq: metric Ansatz} and the dilaton $\phi$ should satisfy simplify to
\begin{equation}\label{eq: diffeq phi}
    \begin{aligned}
        & \dot{\phi}= - \frac{2}{\pi}e^{\phi} \\
        & \dot{G_1}-\frac{1}{18}e^{2G_3-G_1-G_2}-\frac{1}{2}e^{G_2-G_1}+\frac{1}{2}e^{G_1-G_2}=0 \\
        & \dot{G_2}-\frac{1}{18}e^{2G_3-G_1-G_2}+\frac{1}{2}e^{G_2-G_1}-\frac{1}{2}e^{G_1-G_2}=0 \\
        & \dot{G_3} + \frac{1}{9}e^{2G_3-G_1-G_2}-e^{G_2-G_1}-\frac{1}{\pi}e^{\phi}=0  \ , 
    \end{aligned}
\end{equation}
where $\cdot$ refers to a derivative with respect to $\tau$, a convenient radial coordinate defined from $r$ as
\begin{equation}\label{eq: definition of tau}
    d\tau=3e^{-G_3}dr \ .
\end{equation}
The dilaton equation can be easily integrated and gives 
\begin{equation}\label{eq: dilaton solution}
    e^\phi=\frac{\pi}{2}\frac{1}{\tau-\tau_0} \ . 
\end{equation}
For this solution to make sense, we need $\tau>\tau_0$. The dilaton diverges at $\tau=\tau_0$, which we identify as the scale where a duality wall arises in the dual gauge theory. Hence $\tau_0$ represents a natural IR cut-off  for the validity of the solution.\footnote{In \eqref{eq: dilaton solution}, a second integration constant $\phi_0$ has been reabsorbed in the definition of $\tau_0$.}

In order to simplify the remaining equations, let us define the following quantities
\begin{equation}
    \tilde{A}= 3(G_1+G_2) \quad ,\quad \tilde{B}= G_2-G_1 \quad , \quad \tilde{C}= G_1+G_2+G_3 \ ,  
\end{equation}
in terms of which the system of equations becomes  
\begin{equation}
    \begin{aligned}
        & \dot{\tilde{B}}+e^{\tilde{B}}-e^{-\tilde{B}}=0\ , \\
        &  \dot{\tilde{C}}-e^{\tilde{B}}-\frac{1}{\pi}e^\phi=0 \ ,\\
        & \dot{\tilde{A}}-\frac{1}{3}e^{2\tilde{C}-\tilde{A}}=0 \ .
    \end{aligned}
\end{equation}
This can be further simplified by defining
\begin{equation}
    A= e^{\tilde{A}}, \qquad B= e^{\tilde{B}}, \qquad
    C= e^{\tilde{C}} \ , 
\end{equation}
in terms of which we get
\begin{equation}
    \begin{aligned}
        & \dot{B} = 1-B^2\ , \\
         & \dot{\tilde{C}}=B-\frac{1}{2(\tau_0-\tau)}\ ,\\
        & \dot{A}= \frac{1}{3} C^2\ ,
    \end{aligned}
\end{equation}
where \eqref{eq: dilaton solution} has been used in the second equation. This system can now be easily solved. The first equation gives 
\begin{equation}
    \log{\abs{\frac{B+1}{B-1}}}=2(\tau - \tau_c) \ , 
\end{equation}
where $\tau_c$ is a new integration constant. This gives two solutions
\begin{equation}
    \label{eq: solution for B}
    B_1=\coth{(\tau-\tau_c)}\ , \qquad B_2=\tanh{(\tau-\tau_c)} \ , 
\end{equation}
both valid for $\tau>\tau_c$. Note that choosing $B=B_1$ amounts to taking $G_2>G_1$, which, as explained in the section \ref{sec: IR limit of SUGRA}, implies that the $S^3$ shrinks slower than the $S^2$ at $\tau\sim\tau_c$. Since this is the expected behaviour for a supergravity dual to a confining gauge theory, we take $B=B_1$ from now on. Moreover, since in the dual gauge theory the duality wall is always screened by confinement, from now on we will set $\tau \ge \tau_c > \tau_0$. We can now proceed to solve the second equation, substituting the solution we just found 
\begin{equation}
    \dot{\tilde{C}}=\coth{(\tau-\tau_c)}-\frac{1}{2(\tau_0-\tau)}\ . 
\end{equation}
This is easily integrated giving 
\begin{equation}
    \tilde{C}=\log{\sinh{(\tau-\tau_c)}}+\frac{1}{2}\log{(\tau-\tau_0)}+c \ , 
\end{equation}
where $c$ is another integration constant. The third equation then becomes 
\begin{equation}
    \dot{A}= \frac{1}{3} \mu^4 (\tau-\tau_0)\sinh^2{(\tau-\tau_c)}=\frac{1}{6}\mu^4
(\tau-\tau_0)(\cosh{(2(\tau-\tau_c))}-1) \ , 
\end{equation}
where we have set $\mu^2=e^c$. Once again, a simple integral gives 
\begin{equation}
    A=\frac{1}{24}\mu^4 {\tilde{\Lambda}}^3(\tau)\sinh^3{(\tau-\tau_c)} \ , 
\end{equation}
where, in analogy with \cite{Benini:2007gx}, we have defined
\begin{equation}\label{tildelambda}
    \tilde{\Lambda}(\tau)=\frac{[2(\sinh{(
    2(\tau-\tau_c))}-\tau)(\tau-\tau_0)-\cosh{(2(\tau-\tau_c))}+2\tau\tau_0+1-4\tau_c\tau_0+2\tau_c^2+K]^\frac{1}{3}}{\sinh{(\tau-\tau_c)}} \, ,
\end{equation}
with $K$ yet another integration constant.  

\begin{figure}[ht]
\centering
\begin{tikzpicture}[baseline={(image.south)}]
  \node[anchor=south west, inner sep=0] (image) at (0,0) {
    \includegraphics[width=0.65\textwidth]{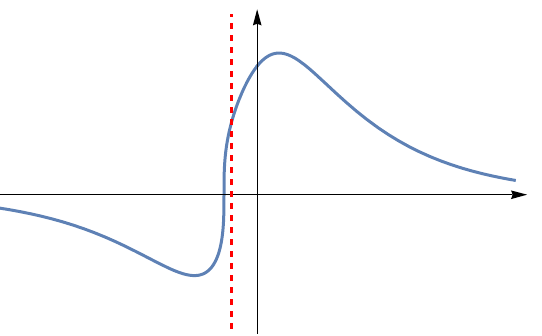}
  };
  \node at (10.8,2.6) {$\tau$};
  \node at (5.8, 6.5) {$\tilde{\Lambda}(\tau)$};
\end{tikzpicture}
      \caption{$\tilde{\Lambda}(\tau)$, when regularity is imposed at $\tau=\tau_c$. The red dashed line is $\tau=\tau_0$.}
      \label{fig: Lambda plot}
  \end{figure}

The physics of the model can help us fix some of the integration constants. In particular, if our solution is to describe confinement in a similar fashion to the KS solution, it needs to end at $\tau=\tau_c$. So, let us look at the asymptotic behavior of $\tilde{\Lambda}$ around this point. The expansion is
\begin{equation}
\label{eq: expansion Lambda}
    \tilde{\Lambda}(\tau)=
    \begin{cases}
         \frac{\sqrt[3]K}{\tau-\tau_c} &\qquad K\neq0 \\
        \frac{2}{\sqrt[3]3}(\tau_c-\tau_0)^\frac{1}{3}+\frac{1}{2\sqrt[3]{3}}(\tau_c-\tau_0)^{-\frac{2}{3}} (\tau-\tau_c)&\qquad K=0, \quad \tau_c> \tau_0 \\
     \sqrt[3]2 (\tau-\tau_c)^\frac{1}{3} &\qquad K=0, \quad \tau_c=\tau_0
    \end{cases}
\end{equation}
The function $\tilde{\Lambda}$ is smooth at $\tau=\tau_c$ only for $K=0$ and $\tau_c>\tau_0$.\footnote{ When $\tau_c = \tau_0$, the dilaton singularity is part of the solution and it can be checked that both the $S^2$ and the $S^3$ shrink to a point at $\tau=\tau_0$, which means that the solution does not describe a confining vacuum. Still, the $F_3$ flux remains finite at the tip and, for large $\tau$, the solution still describes a duality cascade, similarly to what happens in the Klebanov-Tseytlin background. We will not investigate this case further, as it is not clear to what field theory RG flow it could correspond to.} Imposing such regularity, it has the behavior depicted in figure \ref{fig: Lambda plot}. 
Putting all together and substituting back the original functions $G_i$, we get
\begin{equation}
    \begin{aligned}
        & e^{G_2-G_1}=\coth{(\tau-\tau_c)}\ , \\
        & e^{G_1+G_2+G_3}=\mu^2 \sinh{(\tau-\tau_c)} \sqrt{\tau-\tau_0} \ ,\\
        & e^{G_1+G_2}=\left(\frac{1}{24}\right)^\frac{1}{3} \mu^\frac{4}{3} \tilde{\Lambda}(\tau) \sinh{(\tau-\tau_c)} \ , 
    \end{aligned}
\end{equation}
and finally  
\begin{equation}
    \label{eq: solution deformed conifold}
    \begin{aligned}
        & e^{2G_1}= \frac{1}{4} \mu^\frac{4}{3} \tilde{\Lambda}(\tau) \frac{\sinh^2{(\tau-\tau_c)}}{\cosh{(\tau-\tau_c)}}\ , \\
        & e^{2G_2} = \frac{1}{4} \mu^\frac{4}{3} \tilde{\Lambda}(\tau) \cosh{(\tau-\tau_c)}\ , \\
        & e^{2G_3}= 6 \mu^{\frac{4}{3}} \frac{\tau-\tau_0}{\tilde{\Lambda}^2(\tau)}\ ,
    \end{aligned} 
\end{equation}
where $\tilde{\Lambda}(\tau)$ is the expression \eqref{tildelambda} with $K=0$
and we have redefined $\mu$ to absorb some unpleasant numerical factors. 

Finally, we need to determine the functions $f$, $k$ and $F$. They satisfy the following system of differential equations 
\begin{equation} \label{eq: fluxes equation}
\begin{split}
\dot{f} &= e^{\phi} \biggl(1 - F - \frac{2}{\pi} k \biggr) \; \tanh^2{\left(\frac{\tau-\tau_c}{2}\right)} \ , \\
\dot{k} &= e^{\phi} \biggl(F - \frac{2}{\pi} f \biggr) \; \coth^{2} {\left(\frac{\tau-\tau_c}{2}\right)} \ ,\\
\dot{F} &= \frac{1}{2} e^{- \phi} (k - f)\ ,
\end{split} 
\end{equation}
whose solution reads exactly as in \cite{Benini:2007gx}
\begin{equation}\label{eq: solution of fluxes equation}
    \begin{split}
        e^{-\phi} f &= \frac{(\tau-\tau_c)\coth{(\tau-\tau_c)} - 1}{2 \sinh {(\tau-\tau_c)}}(\cosh{(\tau-\tau_c)}- 1)\ ,\\
        e^{-\phi} k &= \frac{(\tau-\tau_c) \coth{(\tau-\tau_c)} - 1}{2\sinh{(\tau-\tau_c)}}(\cosh{(\tau-\tau_c)} + 1) \ ,\\
        F &= \frac{\sinh{(\tau-\tau_c)} - (\tau-\tau_c)}{2 \sinh{(\tau-\tau_c)}} \ . 
    \end{split}
\end{equation}
As we have seen, there are two physically relevant integration constants, $\tau_c$ and $\tau_0$, which correspond in the dual gauge theory to the confinement scale and to the one at which a duality wall would be, respectively. Since $\tau$ is defined in \eqref{eq: definition of tau} up to a shift, one can choose, for example, $\tau_c=0$ and $\tau_0<0$, as we have done in the main text.

\bibliographystyle{utphys}
\bibliography{draft_2}

\end{document}